\documentclass[a4paper,10pt]{article}

\usepackage{float}
\usepackage{amssymb,amsmath}
\usepackage{graphicx}
\usepackage{subcaption}
\usepackage{gensymb}
\graphicspath{{figs/}}
\usepackage[colorinlistoftodos]{todonotes}
\usepackage{xcolor}
\usepackage{multirow}
\usepackage{xfrac}
\usepackage[table]{xcolor}
\usepackage[normalem]{ulem}
\usepackage{tikz}
\usepackage[section]{placeins}
\usepackage{siunitx}
\usepackage{authblk}
\usepackage{natbib}
\usepackage{booktabs}
\usepackage{tabularx}
\usepackage{textcomp}
\usepackage{siunitx}
\DeclareSIUnit{\micro}{\text{\textmu}}
\usepackage{hyperref}
\hypersetup{
	colorlinks,
	citecolor=blue,
	linkcolor=blue,
	urlcolor=blue
}
\usepackage{xcolor}
\usepackage{setspace}
\usepackage[margin=2.5cm]{geometry}

\title{SnapPINN: Pressure and Energy Dissipation Reconstruction from a Sparse and Noisy Velocity Snapshot}
\author[1*]{Robin Barta}
\author[2]{Christian Bauer}
\author[1]{Gholamhossein Bagheri}
\affil[1]{\small{Laboratory for Fluid Physics Pattern Formation and Biocomplexity, Max Planck Institute for Dynamics and Self-Organization 37077 Göttingen, Germany}}
\affil[2]{\small{Institute of Aerodynamics and Flow Technology, German Aerospace Center (DLR), 37073 G\"{o}ttingen, Germany}}
\affil[*]{corresponding author: robin.barta@ds.mpg.de}
\date{\today}

\begin{document}
\onehalfspacing
\maketitle
\begin{abstract}
Reconstructing pressure and turbulence quantities from experimental velocity measurements is challenging, especially without time-resolved data. Furthermore, limitations such as low seeding density, finite resolution, and measurement noise severely hinder the reconstruction of accurate flow fields. 
We introduce SnapPINN, a two-stage physics-informed neural network (PINN) that successfully reconstructs 3D velocity, their spatial gradients, pressure fields and estimates turbulent kinetic energy dissipation from a single snapshot of sparse, noisy velocity data. Evaluated here on 3D DNS turbulent pipe flow data, SnapPINN uses a sine-activated architecture with sequentially trained, decoupled velocity and pressure sub-networks. In stage 1, the velocity network fits particle data while enforcing incompressibility, serving as a physically consistent smoothing operator that regularises velocity gradients against noise. In stage 2, the velocity network is frozen, and the pressure network is trained using the pressure Poisson equation and the pretrained velocity gradients. We systematically map reconstruction performance of SnapPINN across 100 test cases to mimic challenging experimental, such as adding significant position noise, linearization of velocity field and seeding sparsity as low as $0.07\%$ of the fully resolved DNS grid. Quantitatively, bulk velocity was reconstructed within $0.5\%$, while errors remained below $50\%$ for the gradient-sensitive energy dissipation rate and within $4$--$24\%$ for the a~posteriori inferred $\mathrm{Re}_{\tau}$, even under extremely sparse and noisy conditions. 
Finally, we establish a practical reliability map that shows which experimental conditions are likely to yield reliable SnapPINN reconstructions in the absence of ground truth. 
\end{abstract}
\textbf{Keywords:} Physics-informed neural networks, Turbulent flows, Sparse measurements, Measurement noise, Flow reconstruction, Two-frame flow measurements, PTV, PIV
\section{Introduction}
\label{sec:Introduction}
Physics-informed neural networks (PINNs) \citep{karniadakis2021physics} are a powerful class of machine learning models that optimise their weights by simultaneously fitting observational data and strictly enforcing known physical laws via the network's loss function. In fluid mechanics, they have emerged as a promising approach to reconstruct challenging flow quantities—such as pressure \citep{shin2025robust, clark2023reconstructing} and temperature \citep{mommert2024,toscano2024pinns,Barta_2025,volk2025pinn}—from sparse velocity measurements by embedding the governing equations of fluid motion directly into the training process. 
Established flow reconstruction and data-assimilation methods have achieved tremendous success in fluid mechanics. Approaches based on physical data integration \citep{van2013piv,van2019pressure,dontu2026iterative} and numerical simulations \citep{bauer2022assimilation1,bauer2022assimilation2} provide highly accurate flow field characterisations when supplied with sufficiently resolved input data. Recently, novel meshless data assimilation techniques based on radial basis functions \citep{sperotto2022meshless,sperotto2024meshless,lawson2026physics} have further advanced the field. Physics-Informed Neural Networks (PINNs) share this versatile meshless nature but employ a deep learning approach, which has proven exceptionally robust when dealing with spatially sparse or otherwise imperfect data. State-of-the-art PINN architectures typically leverage time-resolved velocity fields—such as those obtained from time-resolved Particle Image Velocimetry (PIV) \citep{Adrian2011,Raffel2018} or time-resolved Particle Tracking Velocitmetry \citep{schanz2016shake,herzog2021probabilistic,barta2023pro,le2026vicctor}—to accurately capture flow dynamics by exploiting information contained in the temporal acceleration term \cite{duan2026physics}. The present work instead addresses the experimentally important setting of two-frame Particle Tracking Velocimetry (PTV) \citep{maas1993particle,Elsinga2006}, where only a single instantaneous velocity snapshot is available. This measurement paradigm is particularly relevant for high-speed turbulent flows and for experiments conducted under practical constraints, such as field measurements or microscale flows, where acquiring time-resolved velocity fields remains prohibitively challenging, e.g. see \cite{thiede2025holotrack}. Furthermore, any experimental measurements is subject to measurement errors and often has limited spatial resolution due to low local seeding density, particularly near high-shear regions. Despite the growing success of PINNs, there remains a limited understanding of the conditions under which they can accurately estimate velocity gradients, turbulent energy dissipation rates, and reconstruct pressure fields in non-thermal flows from a single instantaneous velocity snapshot. Equally important is understanding the extent to which PINNs can compensate for data sparsity and experimental imperfections, and identifying the practical considerations that experimentalists should account for when designing experiments or applying PINNs to maximise the retrieval of flow information. This study addresses these open questions through a systematic investigation of the performance and limitations of PINNs under experimentally relevant conditions. \\
We present a robust and efficient architecture for reconstructing pressure and energy dissipation rates from a single snapshot of three-dimensional velocity data obtained from Direct Numerical Simulation (DNS) of turbulent pipe flow. The method is assessed over a range of data sparsity levels and simulated measurement noise representative of typical experimental conditions. In this work, we introduce the snapshot PINN (SnapPINN) operating at a single velocity timestep. SnapPINN consists of a sine-activated multi-layer perceptron architecture that combines decoupled velocity and pressure output networks with a two-stage training strategy. In the first stage, the velocity network acts as a physically-consistent smoothing function $(u,v,w)=\text{SnapPINN}_1(x,y,z)$, where $u$, $v$, $w$ are the 3D flow velocity components. SnapPINN is trained using a data loss and a divergence-free (incompressibility) constraint ($\nabla\cdot\vec{u}=0$) to regularise velocity gradients and mitigate measurement noise. Thus, the first stage serves as a physical smoothing pretraining step \citep{Wang2016}. In the second stage, the velocity network is frozen, and the pressure network is trained via the pressure Poisson Equation (PPE) using the gradients provided by the frozen velocity network to solve the pressure $p=\text{SnapPINN}_2(x,y,z,u,v,w)$. This transition from a smoothing data fit to a physics-informed reconstruction fills a critical gap in the literature, where single-snapshot performance has previously been missing using PINNs. We evaluate this method using DNS data from a turbulent pipe flow with a friction Reynolds number $\mathrm{Re}_\tau=180$ to provide a reliable ground truth. This flow regime was chosen for its complexity, characterised by high wall-normal gradients, anisotropy, and distinct velocity magnitudes that challenge SnapPINN in a rigorous environment, as pipe flows are known to be extremely challenging to measure experimentally at full scale prior to recent advanced efforts \citep{kuchler2024lagrangian,schanz2026volumetric}. Our analysis systematically evaluates the sensitivity of SnapPINN to spatial sparsity and measurement noise. Spatial sparsity is varied from $0.07\%$ to $17.59\%$ of the DNS grid points, ranging from a extremely dilute to a moderately sparse case. Measurement noise is independently introduced into both particle position and velocity: positional white noise is applied over a range of $5\%$ to $100\%$ of the Kolmogorov length scale \cite{kolmogorov41}, while a linearisation error is introduced to the velocity by increasing the time interval between consecutive steps from $12.5\%$ to $200\%$ of the Kolmogorov time scale \cite{kolmogorov41}. \\
The paper is organised as follows. Section \ref{Dataset} introduces the turbulent pipe flow datasets used for evaluation and discusses the incorporation of measurement noise into the data. Section \ref{Methodology} details the methodology of the two-staged snapshot PINN, which will be called SnapPINN. In section \ref{Results}, SnapPINN is validated against varying seeding densities and noise levels. Finally, section \ref{Conclusion} provides concluding remarks.

%
\section{Dataset}\label{Dataset}
\subsection{Turbulent pipe flow}
\begin{figure}[H]
	\centering
    \includegraphics[width=\linewidth]{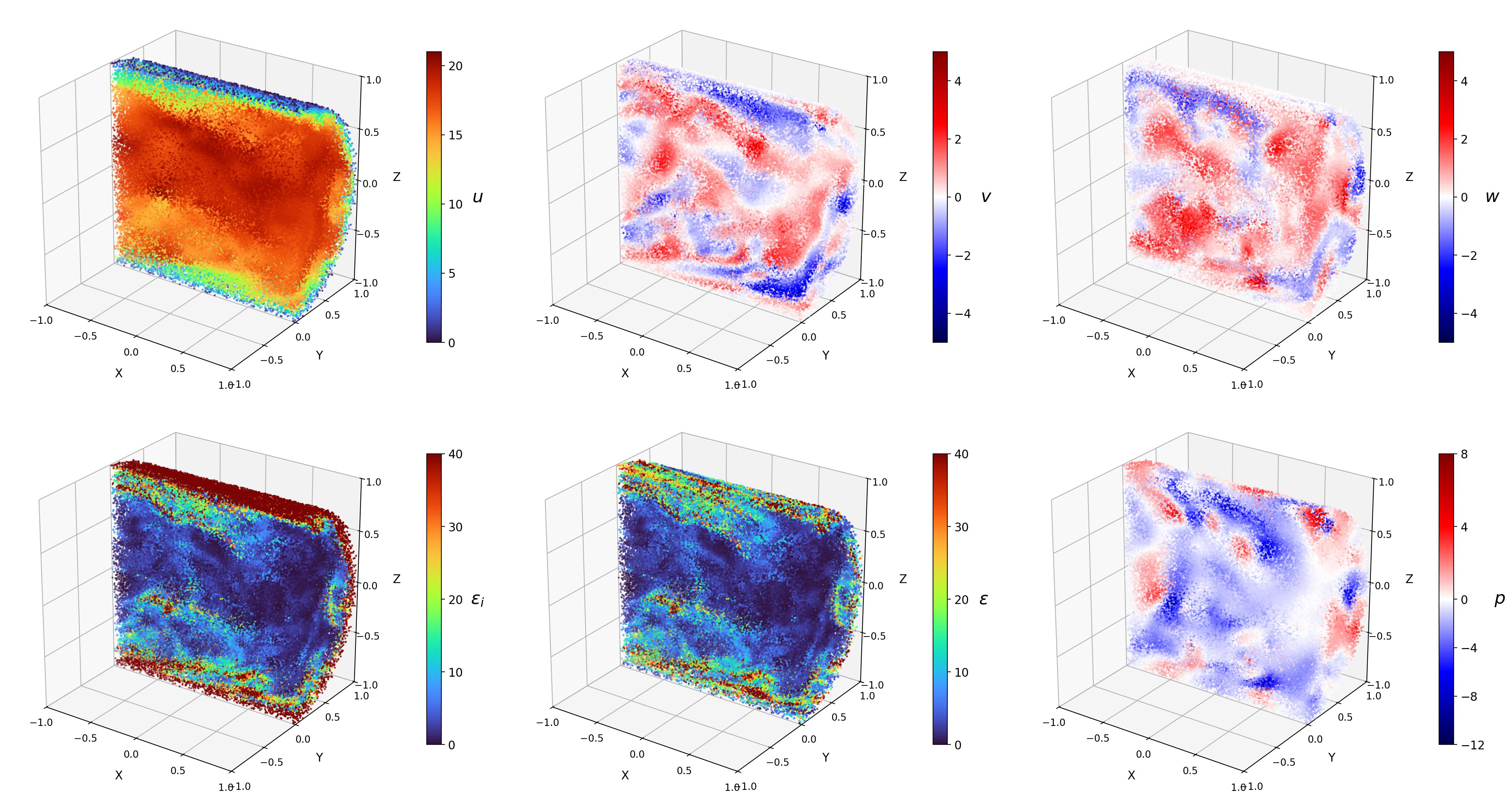}
	\caption{Lagrangian particle fields $u,v,w,\varepsilon_{i},\varepsilon,p$ of an the instantaneous flow field with $N=280000$ data points shown for $Y>0$. The shown time step serves as the ground truth dataset for SnapPINN evaluation.}
	\label{fig:Pipeflow}
\end{figure}

To rigorously assess the performance of SnapPINN, we have considered turbulent pipe flow, which is wall-bounded, characterised by extremely large gradients near the wall, strong anisotropy, and a wide range of interacting spatial and temporal scales. In particular, the different velocity components exhibit markedly different magnitudes and fluctuation intensities, making turbulent pipe flow a demanding benchmark for evaluating the performance of SnapPINN. These characteristics also pose considerable challenges for experimentalists, requiring careful trade-offs in the choice of spatial and temporal resolution to accurately characterise the flow across regions with vastly different flow scales.\\
The governing equations for the pressure-driven incompressible flow of a Newtonian fluid in a smooth pipe are the incompressible Navier-Stokes equations in the following dimensionless form
\begin{align}
\frac{\partial \vec u}{\partial t} + \vec u \cdot \nabla \vec u &= - \nabla p +\frac{1}{\mathrm{Re}_\tau}\Delta \vec u, \label{eq:mom}  \\ 
\nabla \cdot \vec u &= 0,
\label{eq:mass}
\end{align}
where $\mathrm{Re}_\tau=\hat{u}_\tau \hat{R} / \hat{\nu}=180$ is the friction Reynolds number, based on the pipe radius $\hat{R}$.
Here, $\vec{u}$ is the velocity vector normalised by $\hat{u}_\tau$ and $p$ is the pressure normalised by $\hat{\rho}\hat{u}_\tau^2$.
The spatial coordinates $\vec{x}$ are normalised by the pipe radius $\hat{R}$, and time $t$ is normalised by $\hat{R}/\hat{u}_{\tau}$. 
Here, $\hat{u}_{\tau}$, $\hat{R}$, $\hat{\nu}$, and $\hat{\rho}$ denote the dimensional friction velocity, pipe radius, kinematic viscosity, and density, respectively. \\
The present data set is obtained from a DNS, where equation~(\ref{eq:mom}) is integrated in time using a fourth-order finite volume method based on a leapfrog-Euler time integration scheme~\citep{Feldmann2012,Bauer2017,Bauer2019}. 
The flow geometry consists of a smooth circular pipe with length $L=7$ and radius $R=1$. The computational domain is discretised on a staggered grid in a cylindrical coordinate system, where $x$ denotes the axial direction, $\varphi$ the azimuthal direction, and $r$ the radial coordinate. The grid resolution is $N_x \times N_\varphi \times N_r = 256 \times 256 \times 84$.
The grid spacing in axial ($x$) and azimuthal ($\varphi$) directions is equidistant. In contrast, the radial direction ($r$) is non-uniformly discretised, with refinement toward the wall implemented using a hyperbolic tangent stretching function. This allows the simulation to accurately resolve the steep velocity gradients characteristic of wall-bounded turbulent flows.
In viscous or \textit{wall units}, the grid spacings are $\Delta x^+=4.9$, $R^+\Delta \varphi=4.4$, $\Delta r^+_{min}=0.31$, and $\Delta r^+_{max}=4.4$.
Here, normalisation in wall units is denoted by the superscript $+$. The corresponding characteristic scales are the friction velocity $\hat{u}_\tau$ and the viscous length scale $\hat\delta_\nu = \hat\nu / \hat{u}_\tau$. 
For further details on the numerical method the reader is referred to~\cite{Bauer2017}.
The relevant parameters of the ground truth DNS case are summarised in Table~\ref{tab:cases}.
\begin{table}[H] 
\centering
\caption{Ground truth simulation case. $Re_\tau=\hat{u}_\tau \hat{R}/\hat{\nu}$, friction Reynolds number; $\mathrm{Re}_{\text{bulk}}=\hat{U}_{\text{bulk}}\hat{D}/\hat{\nu}$, bulk Reynolds number ($\hat{D}=2\hat{R}$); $U_{\text{bulk}}$, bulk velocity in wall units; $L$, pipe length in radii; $N_x$, $N_\varphi$ and $N_r$ are the number of grid points in the axial, azimuthal and radial directions respectively; $\Delta z^+$, $R^+\Delta \varphi$, $\Delta r^+_{min}$, and $\Delta r^+_{max}$ are the corresponding grid spacings. The superscript $\cdot^+$ denotes spacings normalised in wall units using the viscous length $\hat{\delta}_\nu = \hat{\nu}/\hat{u}_\tau$, with friction velocity $\hat{u}_\tau$; $\Delta t_{stat}$, time interval for statistical averaging; $\Delta t_{DNS}$, DNS time step.}
\begin{tabularx}{1.00\textwidth}{XXXXcXXXXXcX}
\hline \hline
  $\mathrm{Re}_\tau$ &  $\mathrm{Re}_{\text{bulk}}$ & $U_{\text{bulk}}$ &  $L$   & $N_x\times N_\varphi \times N_r$   & $\Delta x^+$ & $R^+\Delta \varphi$&  $\Delta r^+_{min}$  &$ \Delta r^+_{max}$ & $\Delta t_{stat}$ & $\Delta t_{DNS}$\\ 
  \hline  
  180  &  5285 & 14.72 &  7   & $256 \times 256 \times 84$ & 4.9 & 4.4 & 0.31 & 4.4 & 740 & $2\times 10^{-5}$\\	
   \hline \hline 
\end{tabularx}
\label{tab:cases}
\end{table}
To mimic the experimental acquisition of a single velocity snapshot, the DNS flow is seeded with tracer particles that are subsequently tracked in the same manner as in particle tracking experiments. The validation dataset for SnapPINN is obtained from 
tracer particles, which are initially randomly seeded in the volume of the pipe 
and subsequently tracked in time using first-order integration of the DNS velocity field at each DNS timestep. 
This particle-based sampling provides an unstructured set of training points for SnapPINN, and we focus only on particles in the observation volume $-R \le x \le R$. 
Along these Lagrangian trajectories, the velocity components $u$, $v$, and $w$, the pressure $p$ as well as the pressure gradients $\partial p/\partial x_i$, and the instantaneous dissipation rate of turbulent kinetic energy are sampled from the DNS and transformed to Cartesian coordinates. 
Specifically, the instantaneous dissipation rate is defined as
\begin{align}
    \varepsilon_{i} &= \frac{2}{\text{Re}_{\tau}} \text{Tr}\left(\textbf{S}\, \textbf{S}\right) 
\end{align}
where $\textbf{S}=\frac{1}{2}\left(\nabla \vec u + \nabla\vec u^{\,T}\right)$ is the instantaneous strain-rate tensor, whereas the instantaneous turbulent kinetic energy dissipation $\varepsilon$ is defined by:
\begin{equation}
\varepsilon = \frac{2}{\text{Re}_{\tau}} \text{Tr}\left(\textbf{S}^\prime \textbf{S}^\prime\right) \,,
\end{equation}
with a fluctuating part of the strain tensor $\textbf{S}^\prime$ given the Reynolds decomposition: $\textbf{S} = \bar{\textbf{S}} + \textbf{S}^\prime$, see e.g.~\cite{pope2000turbulent}. 
The velocity gradients involved in the calculation of $\textbf{S}^\prime$ and $\textbf{S}$ coincide besides the two terms:
\begin{align}
    \partial_y u' &= \partial_y u - \langle\partial_y u\rangle \\
    \partial_z u' &= \partial_z u - \langle\partial_z u\rangle
\end{align}
because of $\langle\partial_y u\rangle$ and $\langle\partial_z u\rangle$ do not average to 0 in a pipe flow. 
Here, angled brackets denote averaging in time as well as over the homogeneous directions $x$ and $\varphi$. 
Note that in homogeneous isotropic turbulence $\varepsilon_{i}=\varepsilon$. Figure~\ref{fig:Pipeflow} shows an instantaneous snapshot of $u$, $v$, $w$, $\varepsilon_{i}$, $\varepsilon$, and $p$ in the positions of the particles after tracking them for $0.03$ dimensionless time units, and this snapshot serves as the ground truth data on which SnapPINN is validated. The smallest scales of turbulence, known as the Kolmogorov scales \citep{kolmogorov41}:
\begin{align}
    \eta = \left(\frac{1}{\varepsilon \,\text{Re}_{\tau}^3}\right)^{1/4}, \quad\tau = \left(\frac{1}{\varepsilon \,\text{Re}_{\tau}}\right)^{1/2},
\end{align}
are defined by the dissipation of turbulent kinetic energy and the friction Reynolds number. In the following, we utilise the Kolmogorov length scale $\eta$ as the characteristic scale to quantify particle sparsity in the datasets across the different turbulent scales present in the pipe flow. 
Since turbulent kinetic energy dissipation $\varepsilon$ is available from DNS data, the local value of $\eta$ can be evaluated at each particle position. This allows us to interpret the local sampling density relative to the relevant mean turbulent length scale of the flow, the bulk Kolmogorov length scale $\eta_\text{bulk}$.
It should be noted that the mean value of $\eta$ is not uniform throughout the domain. 
Due to the wall-bounded nature of turbulent pipe flow, the dissipation rate---and therefore the Kolmogorov scale---varies systematically with the radial position. 
In particular, the enhanced dissipation in the near-wall region leads to a substantially smaller local Kolmogorov length scale, $\eta$, of 0.009, compared with 0.018 at the pipe centerline, placing significantly greater demands on the spatial resolution required to accurately resolve the flow.
Here, the bulk Kolmogorov length scale, computed as the volume average over the pipe cross section, is used to quantify the measurement sparsity in section \ref{Datasets}:
\begin{equation}
\begin{aligned}
 \eta_\text{bulk} =\frac{1}{2\pi R^3 \Delta t_{stat}}\int_0^{\Delta t_{stat}}\int_{-R}^{R}\int_0^{2\pi}\int_0^R \eta (t,x,\varphi,r)\,r \,\,\text{d} r \, \text{d} \varphi \,\text{d} x \, \text{d} t  = 0.0113.
\end{aligned}
\end{equation}
Similar to the bulk Kolmogorov length scale we define the bulk Kolmogorov time scale by:
\begin{equation}
\begin{aligned}
 \tau_\text{bulk}   = \frac{1}{2\pi R^3 \Delta t_{stat}}\int_0^{\Delta t_{stat}}\int_{-R}^{R}\int_0^{2\pi}\int_0^R \tau (t,x,\varphi,r)\,r \,\,\text{d} r \, \text{d} \varphi \, \text{d} x \, \text{d} t = 0.0242.
\end{aligned}
\end{equation}
\subsection{Training \& validation datasets}\label{Datasets}
Two primary challenges, i.e. sparsity and measurement noise, are present in almost every particle measurement in turbulent flows, and our main goal is to validate the reconstruction of precise velocity fields $\vec{u}$, velocity gradients in from of the energy dissipation $\varepsilon$, and pressure fields $p$ under these challenges using SnapPINN. Data sparsity is related to the number of particles within the observation volume, which limits the resolution of the reconstructed spatial structures. On the other hand, measurement noise plays a critical role in understanding how PINNs perform under realistic experimental conditions. 
To investigate this, we model measurement noise using two distinct error sources: (i) additive Gaussian random noise applied to the particle positions, and (ii) a linear approximation error in velocity, arising from position measurements taken at two discrete time steps. \\
The ground truth dataset used to validate SnapPINN is generated from all particles within the spatial domain $-R \le x \le R$ obtained from a single DNS time step, shown in figure \ref{fig:Pipeflow}, and this baseline dataset is subsequently modified using varying sparsity and measurement noise settings.
\subsubsection{Modeling of sparsity} 
The sparsity in each analyzed dataset is characterised by the mean distance between particles $d$ compared to $\eta_\text{bulk}$ of $N$ particles in the observation volume of the pipe $V=\pi R^2 L_0 $ with the pipe radius $R=1$ and length $L_0=2R$:
\begin{align}
    d = \frac{1}{\eta_{\text{bulk}}}\left(\frac{V}{N}\right)^{1/3} = \frac{1}{\eta_{\text{bulk}}}\left(\frac{2\pi}{N}\right)^{1/3}.
\end{align}
Moreover, the sparsity in each dataset is quantified by the fraction $f$ of the number of particles $N$ relative to the number of grid cells in the observation volume used in DNS, $2N_x/7\times N_\varphi \times N_r$:
\begin{align}
    f = \frac{N}{\frac{2}{7}N_x\, N_\varphi \, N_r}=\frac{N}{1569792}.
\end{align}
We investigate 4 different levels of sparsity ranging from moderately sparse to extremely dilute, listed in table \ref{tab:sparsity}. 

\begin{figure}[H]
	\centering
    \includegraphics[width=\linewidth]{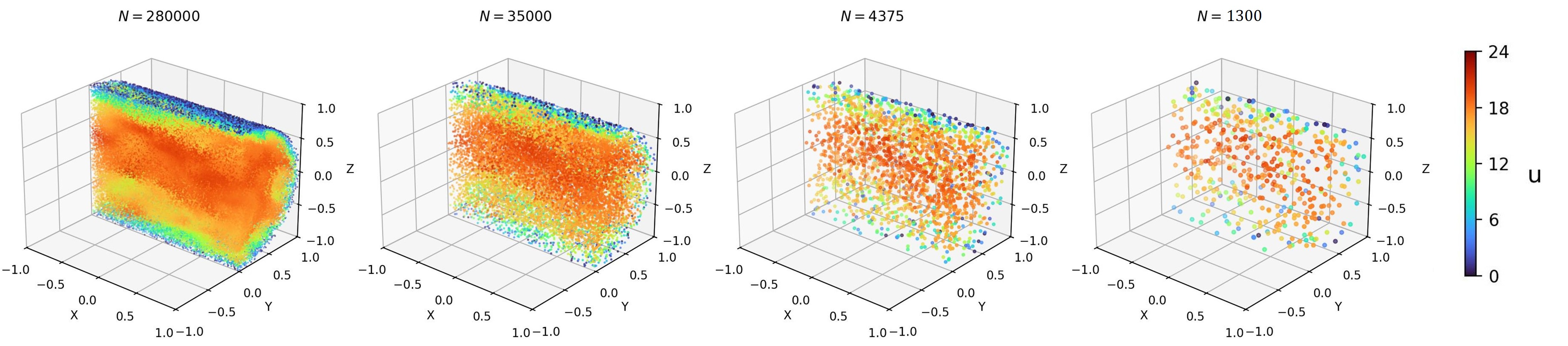}
	\caption{Effect of sparsity on the dataset. Shown are the four levels of sparsity from left to right for particles at $Y>0$ coloured with the $u$ velocity component.}
	\label{fig:sparsity}
\end{figure}

\begin{table}[h]
\centering
\caption{Sparsity parameters of the datasets used to validate SnapPINN. $N$ is the number of particles, $d$ mean inter-particle distance, and $f$ the fraction of particles with respect to DNS grid cells.}
\begin{tabular}{ccc} 
\hline \hline
  $N$ & $d$ [$\eta_{\text{bulk}}$] & $f$ [\%]  \\ 
  \hline  
  280000 & 2.5 & 17.59 \\
  35000 & 5.0 & 2.20 \\		
  4375 & 10.0 & 0.28 \\	
  1300 & 15.0 & 0.07 \\	
   \hline \hline 
\end{tabular}
\label{tab:sparsity}
\end{table}
To systematically evaluate the effect of data sparsity, the number of particles was successively reduced across cases. In each step, a uniform random subset of particles was sampled from the preceding dataset, ensuring that the underlying flow field remained identical while its spatial resolution was incrementally decreased, see figure \ref{fig:sparsity}.
\subsubsection{Modeling of experimental measurement uncertainties}
\begin{figure}[H]
	\centering
    \includegraphics[width=0.85\linewidth]{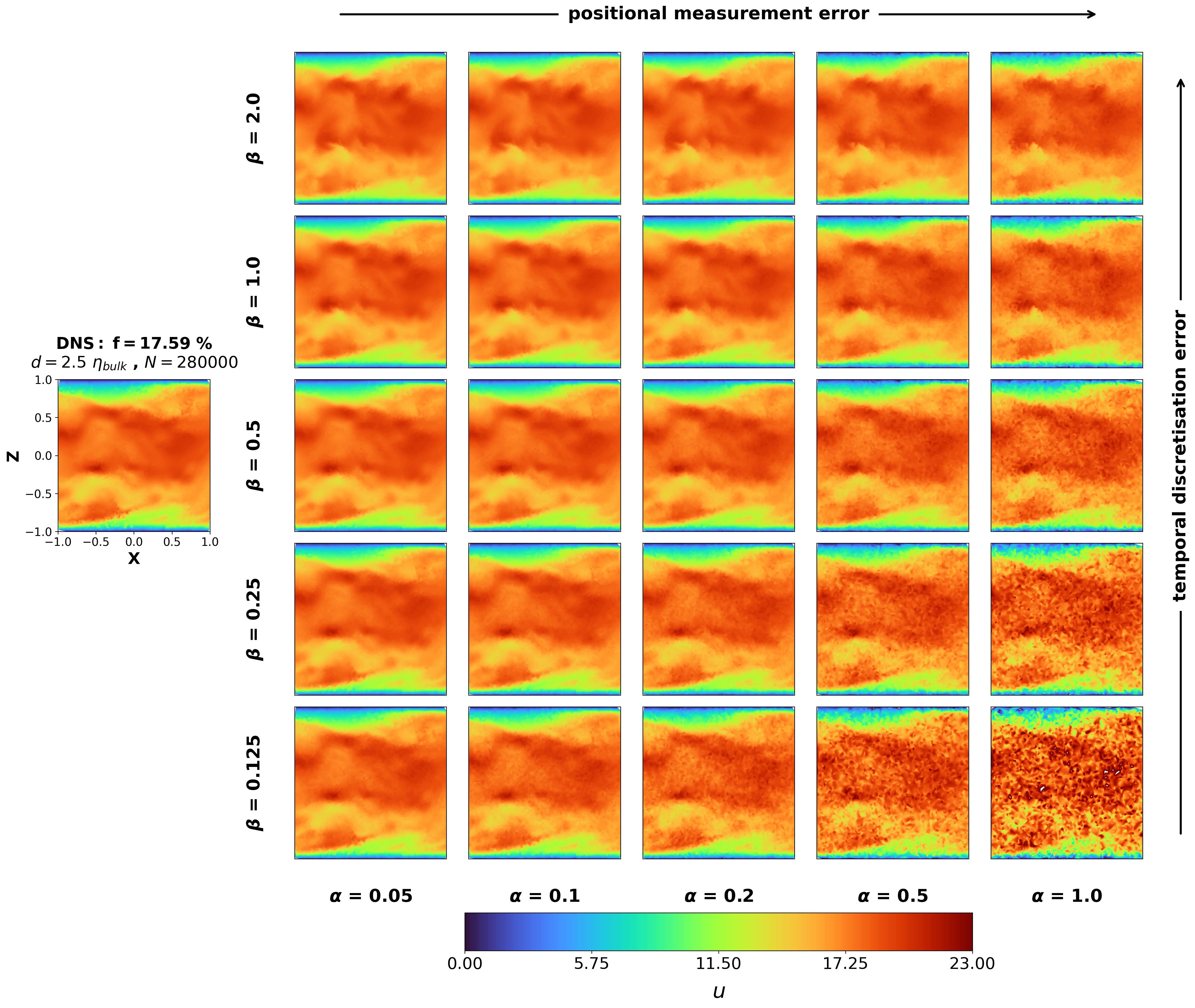}
    \includegraphics[width=0.85\linewidth]{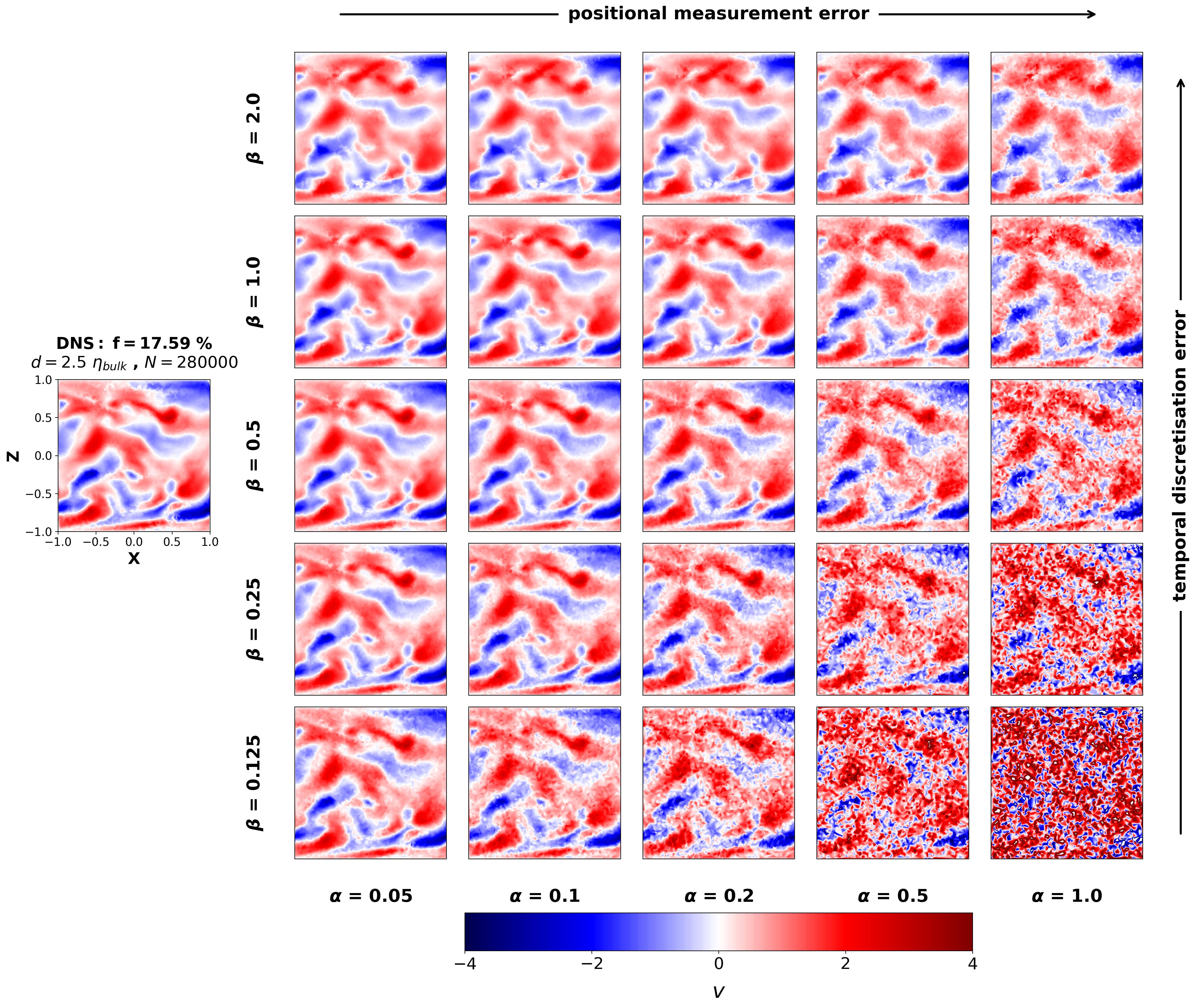}
	\caption{SnapPINN input data: The effect of varying $\alpha$ and $\beta$ on the Lagrangian flow fields $u$ (top) and $v$ (bottom) used as training data for SnapPINN, shown along the axial pipe direction in a slice of thickness $1.5\, \eta_{\text{bulk}}$ around $Y=0$ and plotted as tri-contour from particle positions.}
	\label{fig:noise_overview}
\end{figure}

The use of DNS data is crucial for the validation of SnapPINN, as it gives the exact ground - truth instantaneous $p$ and $\varepsilon$ fields at all particle positions. To bridge the gap between ideal simulations and real-world applications---given that SnapPINN is intended to be used on data obtained from experimental measurements---we model realistic conditions by applying two types of noise inherent to experimental datasets. \\
First, a Gaussian random noise with zero mean and standard deviations $\sigma_{\vec{X}}$ is added to the positions of the particles at each time step considered. Such measurement errors arise from uncertainties in particle position resulting from a combination of limitations associated with the imaging system and measurement principle, including image discretisations due to finite pixel resolution, effective pixel size, and optical imperfections such as defocusing and distortion. Ideally, experiments aimed at characterizing turbulent flows require the positional measurement uncertainty to satisfy $\Delta X \ll \eta_{\text{bulk}}$. Here, we deliberately push beyond this ideal experimental requirement to assess the robustness of SnapPINN under increasingly challenging measurement conditions. To this end, we introduce a parameter $\alpha$ that controls the level of positional noise according to:
\begin{align}
    \sigma_{\vec{X}} = \alpha \,\, \eta_{\text{bulk}} \,, \quad\text{with}\quad\alpha\in\left[\frac{1}{20},\frac{1}{10},\frac{1}{5},\frac{1}{2},1\right].
\end{align}
Second, every Lagrangian two-frame or multi-frame measurement system (e.g. PTV \citep{maas1993particle,schanz2016shake,barta2023pro} or HoloTrack \citep{thiede2025holotrack}) provides measured position at consecutive time steps, and the velocity is estimated to first order by a linear approximation:
\begin{align}
    \vec{u}_{i} \approx \frac{\vec{x}_{i+1}-\vec{x}_{i}}{t_{i+1}-t_{i}}=\frac{\Delta\vec{x}_i}{\Delta t}\,,
\end{align}
between two time steps $t_{i}$ and $t_{i+1}$ with fixed temporal resolution $\Delta t=t_{i+1}-t_{i}$. The temporal resolution is related to the smallest time scales in the flow, e.g. the Kolmogorov time scale or here in case of a pipe flow $\tau_{\text{bulk}}$ and to resolve the smallest flow time scales $\Delta t\ll\tau_{\text{bulk}}$ should be chosen ideally.
Following the same approach as for positional uncertainty, we progressively increase $\Delta t$ beyond the ideal experimental requirement to assess the robustness of SnapPINN to temporal resolution. Specifically, we consider:
\begin{align}
    \Delta t = \beta \,\, \tau_{\text{bulk}} \,, \quad\text{with}\quad\beta\in\left[\frac{1}{8},\frac{1}{4},\frac{1}{2},1,2\right],
\end{align}
For each case, particle velocities are estimated from the finite difference of the noisy particle positions between the DNS time steps $t_0$ and $t_0 + \beta\tau_{\text{bulk}}$, divided by the corresponding time interval $\Delta t$. Consequently, in addition to the imposed spatial sparsity and positional uncertainty, the linear approximation of particle displacement introduces a temporal discretisation error relative to the true DNS velocity field. For all datasets generated with varying $d$, $\alpha$, and $\beta$, any particle perturbed outside the investigated pipe domain by positional noise is discarded. To ensure a consistent spatial distribution across all parameter variations, we apply identical sub-sampling indices to reduce each dataset from 280000 to 1300 particles, i.e. from a particle fraction of $f=17.59\%$ to $f=0.07\%$ of the total DNS grid cells, or varying $d=2.5\eta_{\text{bulk}}$ to $d=15\eta_{\text{bulk}}$. \\
Figure \ref{fig:noise_overview} illustrates how varying $\alpha$ and $\beta$ impacts the Lagrangian velocity fields ($u$ and $v$) for the case of $f=17.59\%$ (i.e. $N=280000$ particles with a mean interparticle distance of $d=2.5\,\eta_{\text{bulk}}$), and the particle fields are shown in an axial slice of thickness $1.5\,\eta_{\text{bulk}}$ along the pipe around $Y=0$. Note that the velocity components $v$ and $w$ exhibit similar spatial structures and added noise errors, as their magnitude is comparable. Therefore, for clarity, we focus on presenting only $v$ and $u$ throughout this study. While increasing $\alpha$ directly amplifies the positional measurement noise, increasing $\beta$ affects the data differently by spatially smearing the flow structures due to increased temporal discretisation of the velocity by linear approximation.
Experimentally measuring all pipe flow velocity components with uniform accuracy is challenging due to their highly disparate magnitudes. Reflecting this limitation, the cross-stream components ($v$ and $v$) in our datasets are six times smaller in magnitude than the streamwise component ($u$). This results in approximately six-fold larger relative errors when the absolute velocity uncertainty is comparable across all components, as is typically the case experimentally. Increasing the time separation between tracking frames increases the measured particle displacement and thereby reduces the relative uncertainty in the estimated velocity; however, larger time separations also increase the temporal discretization error associated with the finite-difference approximation. This introduces an inherent experimental trade-off between these two sources of uncertainty and raises the question of whether PINNs can mitigate their effects during post-processing. In summary, we generate 100 test cases spanning different combinations of sparsity $d$, positional noise $\alpha$, and temporal resolution $\beta$.

\section{Methodology}\label{Methodology}
\subsection{SnapPINN Architecture}
\begin{figure}[H]
	\centering
    \includegraphics[width=\linewidth]{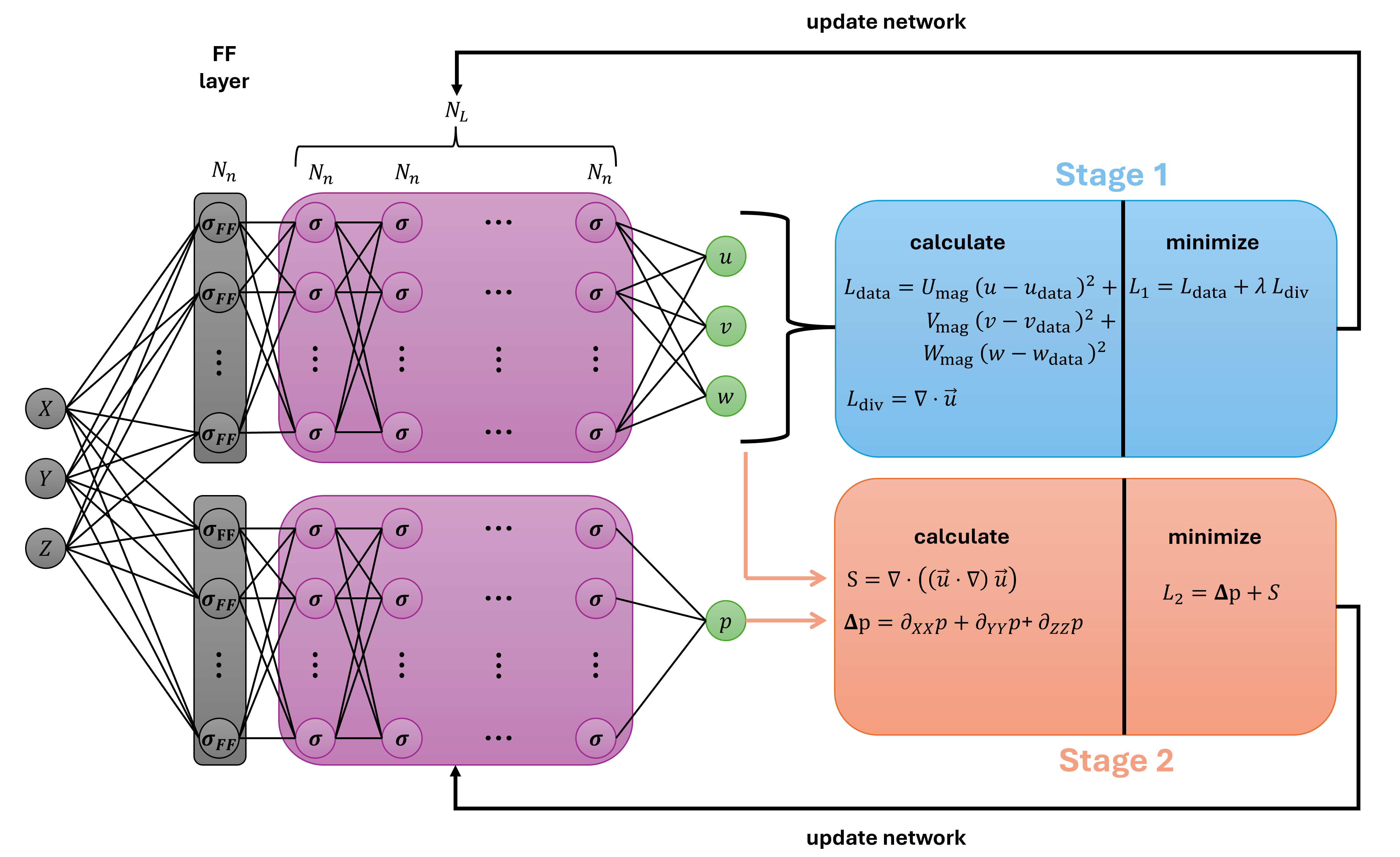}
	\caption{Architecture of the SnapPINN network. The split architecture involves separate velocity and pressure networks which are trained after another in two different stages.}
	\label{fig:PINN}
\end{figure}

The proposed SnapPINN features a decoupled architecture (see Figure \ref{fig:PINN}) consisting of two independent multi-layer perceptrons (MLPs), designed to robustly reconstruct three-dimensional sparse velocity and pressure fields from a single measurement time step. 
As illustrated in the network topology, the architecture separates the reconstruction task into two stages: (stage 1) the velocity network takes the spatial coordinates $(X, Y, Z)$ as input and outputs the three velocity components $(u, v, w)$; (stage 2) the pressure network uses the same spatial input to predict a single scalar value for the pressure $p$. To enable the network to resolve the multi-scale, high-frequency spatial structures inherent in turbulent pipe flow, the first layer of both sub-networks is initialised as a Fourier feature layer \citep{Tancik2020} using the activation $\sigma_{FF}=\sin(2\pi \,B \,\vec{X} + \vec{c})$ with a random normal matrix $B$ and a phase shift $\vec{c}$. This layer embeds the low-dimensional spatial coordinates into a higher-dimensional feature space using random sinusoidal mappings with a fixed phase shift allowing the network to find suitable wavenumbers across different scales much faster. Both the velocity and pressure networks consist of $N_L = 6$ fully connected hidden layers, each with a constant width of $N_N = 300$ neurons. This specific network size was adopted from previous successful PINN implementations to reconstruct quantities in turbulent flow \citep{Barta_2025,mommert2024}, as these parameters provide a proven balance between expressivity and computational efficiency. Generally, the network width $N_N$ allows the model to linearly combine more basis functions, capturing a broader spectrum of turbulent scales. However, making the layers too wide can lead to overfitting at specific collocation points and severe memory bottlenecks, as the memory cost of the second-order automatic differentiation required to express the Laplace operators scales quadratically. On the other hand, increasing the network depth $N_L$ enhances the model's expressivity, enabling it to hierarchically compose the highly non-linear mappings necessary to capture complex spatiotemporal fluid dynamics. This comes at a cost, as calculating higher-order PDE derivatives through many layers exacerbates vanishing and exploding gradients, making deeper networks highly stiff and leaving them with notoriously non-convex loss landscapes \citep{Wang2021,mehlig2021machine}. \\
Furthermore, we utilise the sine activation function, $\sigma(\cdot) = \sin(\cdot)$, throughout the hidden layers following the SIREN architecture \citep{Sitzmann2020}. Sine-activated networks are uniquely suited for turbulent flows; they are resistant to the vanishing gradient problem \citep{Sitzmann2020} and allow the network to rapidly approximate the flow as a superposition of various wavenumbers, which is critical for representing turbulent structures across different scales. To optimise the training process, we initialise the hidden layers of both networks with weights drawn from a Glorot normal distribution and set all biases to zero \citep{Glorot2010}.

\subsection{Training strategy}
The training of SnapPINN is implemented in a Python framework using Tensorflow v.2.16 and Keras v.3.4 \citep{Abadi2016,Chollet2015} using the Adam optimiser \citep{Kingma2014}.
The training strategy relies on the split network architecture discussed above and shown in figure \ref{fig:PINN}. This split structure is essential to reconstruct $\varepsilon$ and $p$ from one time step of velocity data since the optimisation of one time step with both the incompressibility and the pressure Poisson equation in one network is ill posed \citep{duan2026physics}. To overcome this issue, in the first training stage, only the velocity network is trained forcing incompressibility from the data by minimizing the cost function:
\begin{equation}
L_{1} = L_{\text{data}} + \lambda_{\text{div}}\,L_{\text{div}}
\end{equation}
where $\lambda_{\text{div}}$ is the weight of the divergence penalty and 
\begin{align}
L_{\text{data}} =  \frac{1}{N_B}\sum_{i=1}^{N_B}  U_{\text{mag}}(u-u_{\text{data}})^2 
+ V_{\text{mag}}(v-v_{\text{data}})^2
+ W_{\text{mag}}(w-w_{\text{data}})^2\,,
\end{align}
\begin{align}
L_{\text{div}} = \frac{1}{N_B}\sum_{i=1}^{N_B}(\nabla \cdot \vec{u})^2\,.
\end{align}
That way, the PINN reconstruction of the flow acts as a physically consistent smoothing function optimised via a combination of a data fidelity loss and the incompressibility condition leading to physically regularised spatial velocity gradients in the flow. To account for the highly anisotropic nature of pipe flow, where the streamwise velocity dominates the wall-normal and azimuthal components, the data loss contributions are scaled. The weights $U_{\text{scale}}, V_{\text{scale}}$, and $W_{\text{scale}}$ are calculated based on the ratio of the maximum squared velocity magnitude to the maximum squared component magnitude per training batch of size $N_B$.
\begin{align}
    U_{\text{mag}}&= \max_{i \in\{1 \dots N_B\}}\frac{\Vert\vec{u}_{\text{data},i}\Vert^2}{u^2_{\text{data},i}} \\
    V_{\text{mag}}&= \max_{i \in\{1 \dots N_B\}}\frac{\Vert\vec{u}_{\text{data},i}\Vert^2}{v^2_{\text{data},i}} \\
    W_{\text{mag}}&= \max_{i \in\{1 \dots N_B\}}\frac{\Vert\vec{u}_{\text{data},i}\Vert^2}{w^2_{\text{data},i}}
\end{align}
We explicitly do not impose boundary conditions through a separate loss term.
Given the predominantly data-driven optimisation in stage 1, the proposed training strategy is not dependent on explicitly prescribed boundary conditions and can therefore be applied to flows with or without physical boundaries. \\
The training process in stage 1 has only one adjustable variable, which is the weight of the divergence penalty $\lambda_{\text{div}}$, which according to the scaling estimation from \cite{clark2023reconstructingv} should scale with the squared pipe radius, which is $R=1$: 
\begin{align}
    \lambda_{\text{div}} \sim \frac{L_{\text{data}}}{L_{\text{div}}}\sim\frac{U^2}{\partial_xU^2}\sim R^2.
\end{align}
The optimal value of  $\lambda_{\text{div}}$ depends on the noise parameters $\alpha$ and $\beta$,  as well as on the data sparsity $d$.
Therefore, no single value can be expected to be universally optimal across all measurement conditions, and an extensive hyperparameter search is of limited practical value, particularly because fine-tuning $\lambda_{\text{div}}$ yields only marginal improvements in the reconstruction accuracy.
For example, varying $\lambda_{\text{div}}$ from $1$ to $10$ for the pipe flow considered here improved the correlation between the SnapPINN predictions and the DNS reference by 7\%. Values $\lambda_{\text{div}}<1$ led to unphysical solutions for the reconstructed pressure and values above $10$ and below $100$ provided no meaningful improvement. Values above $100$ strongly smoothed the velocity fields. Thus, to be consistent throughout the test cases, we fixed $\lambda_{\text{div}}=10$.
Moreover, $\lambda_{\text{div}}$ controls a trade-off between data fidelity and physical regularisation: excessively large values impose strong smoothing of the velocity field and can consequently blur physically relevant gradients, whereas excessively small values provide insufficient regularisation and may result in unphysical velocity gradients. \\
Dynamic loss balancing is generally preferred, as it aims to promote a more uniform optimisation towards global minima.
The literature suggests using methods such as adaptive loss balancing \citep{Wang2021,Wang2023} and updating the weights dynamically based on a moving average of normalised gradients of the loss function with respect to the network parameters at each epoch. 
However, utilizing this method resulted in an unstable training without converging solutions. 
We hypothesise that this behaviour arises from the substantially more complex nature of the flow considered here compared with the low-dimensional benchmark problems on which these approaches have primarily been demonstrated.
In particular, the pipe flow data considered here is highly turbulent, three-dimensional, anisotropic, and noisy, combined with small batch sizes relative to the total dataset, necessitated by the high GPU memory required for second-order autodifferentiation, which leads to noisy weight updates during training. \\
In stage 2, the weights of the velocity network are completely frozen, and the pressure network is trained independently by minimizing the residual of the pressure Poisson equation
\begin{align}
\label{eq:PPeq}
    \nabla\cdot (\vec{u}\cdot\nabla\vec{u}) = - \Delta p \,,
\end{align}
derived from applying the divergence operator to the Navier-Stokes equation (\ref{eq:mom}). For an incompressible flow, this residual loss is formulated as:
\begin{align}
L_{2} =  \frac{1}{N_B}\sum_{i=1}^{N_B}\left(\Delta p + \nabla (\vec{u}\cdot\nabla \vec{u}) \right)^2 \,.
\end{align}
The source term, $S = \nabla \cdot (\vec{u} \cdot \nabla \vec{u})$, is computed explicitly via the automatic differentiation of the frozen velocity field. Crucially, because the velocity network's parameters are fixed, the computation of $S$ is detached from the active computational graph. This strategic detachment drastically reduces memory overhead and isolates the optimisation of the pressure field. \\
To ensure the PDE residuals are evaluated efficiently throughout the domain, we employ a hybrid collocation point generation strategy. The total number of collocation points per batch is set to $N_{\text{col}} = 20 \times N_{\text{B}}$. Half of these points are sampled uniformly across the cylindrical domain to ensure broad physics enforcement. The remaining half are generated by adding Gaussian noise (with a standard deviation of 5\% of the domain length) to the exact spatial coordinates of the experimental data points. This localised sampling ensures the network heavily penalises unphysical gradient fluctuations precisely where the data is most dense \citep{Lu2021,Nabian2021}. The batch size $N_B$ is adaptively selected: for densely seeded flow fields (e.g., $N=280000$), $N_B$ is maximised to the limit of the GPU VRAM to maintain computational feasibility. For sparser datasets, a batch size of approximately $N/3$ was found to offer the optimal balance between stochastic gradient descent noise and convergence speed. 
This adaptive selection reflects the fact that neither very small nor very large batch sizes are universally optimal, as batch size influences both optimisation dynamics and convergence behaviour \citep{Keskar2017}.
Each dataset is trained as follows: Stage 1 is trained for 1000 epochs with an initial learning rate of $10^{-3}$, while Stage 2 is trained for 3,500 epochs with a reduced initial learning rate of $3 \times 10^{-4}$. The lower initial learning rate for Stage 2 is critical for stable convergence of the training. To avoid introducing additional hyperparameters that would require tuning, we employ the CosineDecay learning rate schedule implemented in tensorflow, which automatically reduces the learning rate during training of each stages to a minimum of $10^{-6}$. This approach scales the learning rate according to a half-cosine function, where the decay is slow at the beginning for rapid exploration and end of training for fine tuning, but steep in the intermediate stages.
Table \ref{tab:time} summarises the network parameters used for training, as well as the computational time per epoch with respect to sparsity $d$, number of particles $N$, and used batch size $N_B$, measured on a V100 Nvidia GPU. 
\begin{table}[H]
\centering
\caption{Network parameters used to train SnapPINN on each dataset. $N_N$ is the number of neurons per layer, $N_L$ number of layers, $N_1$ number of training epochs of stage 1, $N_2$ number of training epochs of stage 2, $N_{\text{col}}$ number of collocation points per batch, $N_{\text{B}}$ batch size, $\text{lr}_1$ initial learning rate of stage 1, $\text{lr}_2$ initial learning rate of stage 2, $\lambda_{\text{div}}$ weight of the divergence-free loss in stage 1, $t_1$ time per training epoch in stage 1 and $t_2$ time per training epoch in stage 2.}
\setlength{\tabcolsep}{5pt}
\begin{tabularx}{\textwidth}{cccccccccccccc}
\hline \hline
  $N_N$ & $N_L$ & $N_1$ & $N_2$ & $N_{\text{col}}$ & $\text{lr}_1$ & $\text{lr}_2$ & $\text{lr}_{\text{end}}$ & $\lambda_{\text{div}}$ & $d$ [$\eta_{\text{bulk}}$] & $N$ & $N_B$ & $t_{\text{1}}$ [s] & $t_{\text{2}}$ [s]  \\ 
  \hline  
  300 & 6 & 1000 & 3500 & 20$N_B$ & $10^{-3}$ & $3\cdot10^{-4}$ & $10^{-6}$ & $10$ & 2.5 & 280000 & 8000 & 1.68 & 4.71 \\
   &  &  &  &  &  &  &  &  & 5.0 & 35000 & 4000 & 0.23 & 0.69\\
   &  &  &  &  &  &  &  &  & 10.0 & 4375 & 1000 & 0.06 & 0.11 \\
   &  &  &  &  &  &  &  &  & 15.0 & 1094 & 350 & 0.04 & 0.06\\
   \hline \hline 
\end{tabularx}
\label{tab:time}
\end{table}
Although SnapPINN requires only a few hyperparameters to set, adjusting the hyperparameters shown in table \ref{tab:time} (left side) besides the divergence loss weight $\lambda_{\text{div}}$, still affects the overall results across the investigated datasets. Identifying the optimal configuration therefore remains a separate challenge in itself. However, empirical tests showed that extensive parameter tuning does not substantially alter the results. Instead, it provides only marginal improvements, with no performance gains exceeding 4\% in PCC accuracy. 
Overall, this reduced hyperparameter space makes SnapPINN straightforward to configure, demonstrating that exhaustive tuning is not necessary to achieve trustworthy results.
\subsection{Evaluation metrics}\label{metrics}
The proposed SnapPINN method is evaluated on each training dataset (with different noise and sparsity values) using four performance metrics. The first metric is the well-known Pearson correlation coefficient (PCC) \citep{schober2018correlation}, and $\text{PCC}_{\zeta}$, for the (instantaneous) flow variables $\zeta\in\left\{u,v,w,\varepsilon,p\right\}$, is defined as:
\begin{align}
    \text{PCC}_{\zeta} = \frac{N \sum_{i=1}^{N} \zeta_i \zeta^*_i - \sum_{i=1}^{N} \zeta_i \sum_{i=1}^{N} \zeta^*_i}{\sqrt{N \sum_{i=1}^{N} \zeta_i^2 - \left(\sum_{i=1}^{N} \zeta_i\right)^2} \sqrt{N \sum_{i=1}^{N} (\zeta^*_i)^2 - \left(\sum_{i=1}^{N} \zeta^*_i\right)^2}}\,.
\end{align}
Properties $\zeta^*$ denote the flow variables from the ground truth reference (DNS), respectively. The PCC assesses the overall structural agreement of the SnapPINN reconstructions with the ground truth. \\
Additionally, we evaluate the capability of SnapPINN to reconstruct integral flow properties by comparing them to the baseline DNS values; namely, the bulk-averaged velocity (\ref{eq:ub}), the bulk-averaged turbulent kinetic energy dissipation (\ref{eq:eb}), and the friction Reynolds number, $\text{Re}_{\tau}$ (\ref{eq:Re}), which is derived from the wall-normal velocity gradient and is related to the normalised wall shear stress. Note that the ground-truth value of $\text{Re}_{\tau}$ is never explicitly provided to SnapPINN during training, nor is it included in the loss functions, because we train with the Pressure Poisson equation \ref{eq:PPeq} instead of the Navier-Stokes equation \ref{eq:mom}. Consequently, evaluating this property after training provides valuable insights into how accurately dimensionless numbers or flow properties can be reconstructed using PINNs, which is an active research question \citep{Mommert2026}. 
\begin{align}\label{eq:ub}
U_{\text{bulk}}  = \frac{1}{2\pi R^3 }\int_{-R}^{R}\int_0^{2\pi}\int_0^R u(x,\varphi,r)\,r \,\,\text{d} r \, \text{d} \varphi \, \text{d} x  
\end{align}
\begin{align}\label{eq:eb}
\varepsilon_\text{bulk}   = \frac{1}{2\pi R^3 }\int_{-R}^{R}\int_0^{2\pi}\int_0^R \varepsilon(x,\varphi,r)\,r \,\,\text{d} r \,\text{d} \varphi \, \text{d} x 
\end{align}
\begin{align}\label{eq:Re}
\text{Re}_{\tau} = -\left.\frac{\partial \langle u\rangle}{\partial r}\right\vert_{r=R}
\end{align}
\section{Results}\label{Results}
All metrics reported in this section are calculated using SnapPINN reconstructions evaluated at the ground truth particle positions directly exported from the DNS. Because the network is trained solely on the noised particle positions characterised by varying $\alpha$ and $\beta$, evaluating at the ground truth locations constitutes an additional task beyond mere denoising. 
\subsection{SnapPINN performance in recovering integral flow quantities}
\begin{figure}[H]
	\centering
    \includegraphics[width=\linewidth]{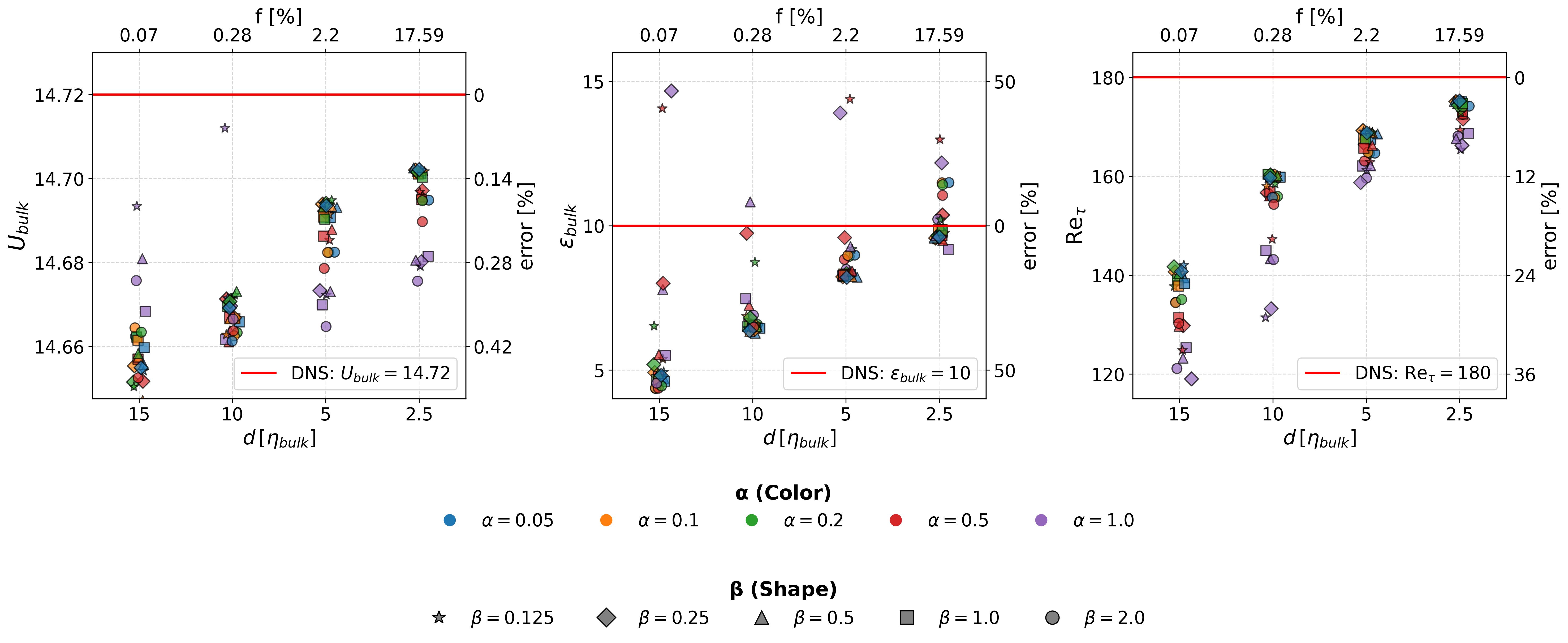}
	\caption{Bulk quantities: instantaneous bulk velocity $U_\text{bulk}$ (left), turbulent dissipation rate $\varepsilon$ (mid) such as the wall quantity $\text{Re}_{\tau}$ (right) obtained from SnapPINN reconstruction on various noised datasets (symbols) at different data sparsity. Red lines indicate the converged DNS ground truth.}
	\label{fig:ubtauw}
\end{figure}

Figure~\ref{fig:ubtauw} summarises the reconstruction quality of the integral flow quantities. Here, we differentiate between bulk-averaged quantities---namely the bulk-averaged velocity $U_\text{bulk}$ and the bulk-averaged turbulent dissipation rate $\varepsilon_\text{bulk}$---and the wall-bounded friction Reynolds number, $\text{Re}_{\tau}$, which, represents a particularly challenging quantity to recover, as its estimation relies on the wall shear stress and therefore requires an accurate reconstruction of the steep velocity gradients in the near-wall region. This sensitivity becomes especially critical for spatially sparse and noisy velocity measurements, where the near-wall flow is inherently difficult to resolve. As it can be seen in Figure~\ref{fig:ubtauw}, across the full parameter study, all three quantities remain closely clustered for variations in $\alpha$ and $\beta$, except for the most extreme and  physically unrealistic noise configurations, namely $\alpha=1$ (highest white noise considered here) and $\beta=2$ (strongest smearing of the velocity fields considered here). This robustness suggests that data noise does not heavily alter integral flow properties due to the inherent physical smoothing nature of PINNs. However, increasing the data fraction $f$ (i.e. increasing number of available particles $N$ or decreasing average distance $d$ between particles) reveals a clear trend: all metrics progressively converge toward the expected DNS values, marked by the solid red lines in each subplot. Specifically, quantifying the relative errors on the secondary vertical axes reveals distinct sensitivities for each integral variable as spatial sparsity increases (i.e., as $f$ decreases to 0.07\% and $d$ increases to $15\eta_{\text{bulk}}$). The bulk velocity $U_\text{bulk}$ proves highly robust to spatial sparsity, maintaining a remarkably low maximum error of less than 0.5\% even for the sparsest configurations. In contrast, the bulk dissipation rate $\varepsilon_\text{bulk}$ is more sensitive to the seeding density. Because $\varepsilon_\text{bulk}$ depends on velocity gradients, accurately fitting the velocity field alone, which suffices for predicting $U_\text{bulk}$, is not sufficient for accurately reconstructing $\varepsilon_\text{bulk}$. Nonetheless, even at $f=2.2\%$ the deviation in estimating $\varepsilon_\text{bulk}$ remains below 20\% for most cases. At extremely sparse cases the deviations in estimating $\varepsilon_\text{bulk}$ grows to nearly 50\%, which nevertheless represents substantial recovery given the limited amount of available velocity information. qSimilarly, the accuracy of the friction Reynolds number $\mathrm{Re}_{\tau}$ decreases with increasing data sparsity, with an underprediction of the DNS reference that grows from approximately $4\%$ at $f=17.59\%$ to about $24\%$ for the most extreme sparsity case ($f=0.07\%$). Notably, $\mathrm{Re}_{\tau}$---a fundamental flow parameter inferred a posteriori from the wall-normal gradient of the reconstructed mean profile---successfully follows the DNS trend at higher particle concentrations without being explicitly prescribed during training. 
\subsection{Practical reliability map for SnapPINN without ground truth}
\begin{figure}[H]
	\centering
    \includegraphics[width=0.96\linewidth]{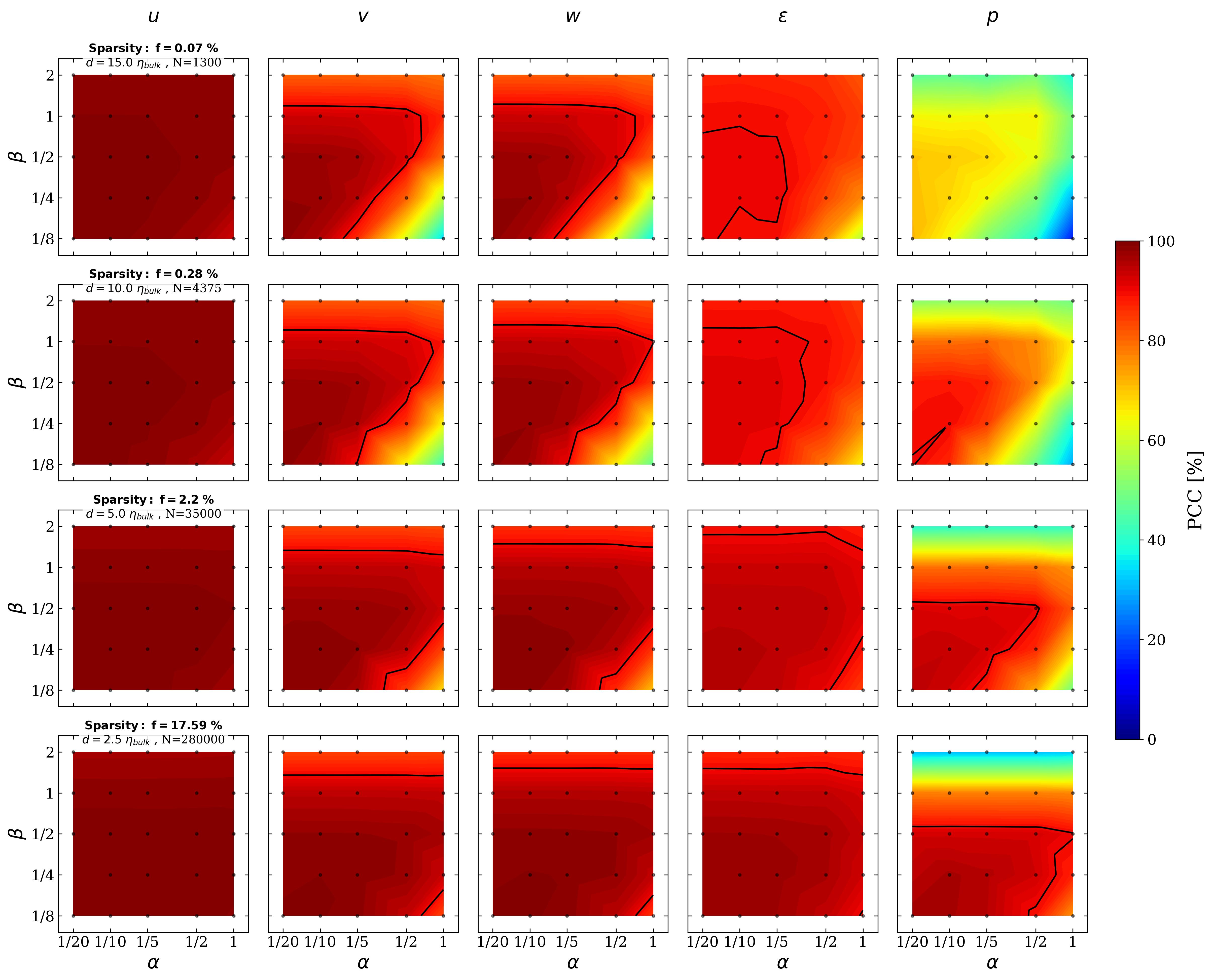}
	\caption{PCC for all flow quantities $u,v,w,\varepsilon$ and $p$ for different sparsity levels $d$ (along the columns) estimated from comparing the ground truth particle fields with the predicted fields using SnapPINN on the datasets noised by different $\alpha$ (position noise) and $\beta$ (discretization noise) values. The black contour line indicates a PCC of 90\%, which indicates a reliable structural agreement of the reconstructed fields with the DNS.}
	\label{fig:pcc}
\end{figure}

In this section, we aim to provide a practical guideline in the form of a lookup map that serves two purposes: (i) to provide a comprehensive overview of the performance of SnapPINN in reconstructing multiple flow quantities across the investigated parameter space of $\alpha$ and $\beta$ at four different sparsity levels $f$, and (ii) to provide practical guidance on the experimental conditions under which SnapPINN is most likely to provide reliable reconstructions when ground-truth data are not available. Although the present map is derived specifically for turbulent pipe flow, this configuration represents a particularly demanding reconstruction problem due to the strong near-wall gradients, pronounced flow anisotropy, and substantially different dynamic ranges of the individual velocity components. We therefore expect the identified performance regimes to provide a useful first-order guideline for a broader range of turbulent flows, with the precise boundaries expected to depend to some extent on the specific flow configuration.
Figure~\ref{fig:pcc} shows the complete parameter space investigated in this study, comprising a total of 100 cases to reconstruct the five flow quantities $u,v,w,\varepsilon$ and $p$ considered here. Each case is coloured according to the corresponding PCC, which can be used to assess qualitatively the structural agreement of reconstructed flow fields using SnapPINN with the ground truth from the DNS. 
As demonstrated in the preceding sections, bulk (integral) flow quantities are recovered with reasonable accuracy across the investigated conditions. Even for the gradient-sensitive bulk energy dissipation rate, deviations remain within approximately $50\%$ at the extreme sparsity level of $f=0.07\%$. Consequently, the bulk flow quantities and characteristic scales required to determine $\alpha$, $\beta$, and $d$ can be estimated at least approximately from the experimental data when not already known, even under substantially non-ideal measurement conditions. An experimentalist can then locate the corresponding operating point within the parameter space of figure~\ref{fig:pcc} and use the results as a practical indication of which flow quantities are likely to be reconstructed reliably by SnapPINN and which should be interpreted with greater caution. As an illustrative example, HoloTrack, a fast two-frame in-situ holographic instrument \citep{thiede2025holotrack} , can be placed within the SnapPINN reliability map using representative cloud measurements. For a Kolmogorov length scale of $\eta \approx 1~\mathrm{mm}$, a Kolmogorov time scale of $\tau \approx 100~\mathrm{ms}$, and droplet number densities of approximately $100$--$1000~\mathrm{cm}^{-3}$, the corresponding measurement conditions are $\alpha \approx 0.1$, $\beta \approx 10^{-3}$, and $f > 20\%$, placing the instrument within a favourable region of the investigated parameter space for SnapPINN reconstruction.\\
As shown in figure~\ref{fig:pcc}, the velocity components in general exhibit consistently high correlation across most $(\alpha,\beta,d)$ combinations, with the streamwise component proving to be the most robust due to its dominating magnitude. In contrast, the cross-stream velocity component and the pressure demonstrate a stronger sensitivity to increased perturbation amplitudes and reduced particle counts. This leads to localised regions of reduced PCC, most profoundly where the dataset is excessively noisy ($\alpha \geq 0.5$) or where the velocity fields are smeared out by the strict linear approximation imposed when $\beta=2$. Remarkably, the SnapPINN is still able to recover the velocity fields in high-noise cases (large $\alpha$) provided the sparsity is not too low. This highlights the strength of the physical interpolation and smoothing capabilities of PINNs when constrained by the divergence-free loss function. The turbulent dissipation rate exhibits an intermediate behaviour: its global correlation remains high for moderate noise levels, but extreme perturbations and extreme sparse sampling reduce the correlation noticeably. Furthermore, for all cases where $\beta=2$ (where the underlying velocity fields are heavily deformed by the linear approximation), no configuration for the flow fields $v,w,\varepsilon,p$ in figure~\ref{fig:pcc} surpasses the 90\% PCC threshold. This is expected, as the structural deformation of the velocity field, most profoundly in $v$ and $w$, has a compounding impact on the gradient-based quantities $\varepsilon$ and $p$, severely degrading the PCC even if the reconstructed fields still appear qualitatively physical, as it is shown in the top rows of each panel in figure \ref{fig:3D_eps} and \ref{fig:3D_p} respectively.
\subsection{SnapPINN performance in recovering instantaneous flow quantities} 
\begin{figure}[H]
	\centering
    \includegraphics[width=0.89\linewidth]{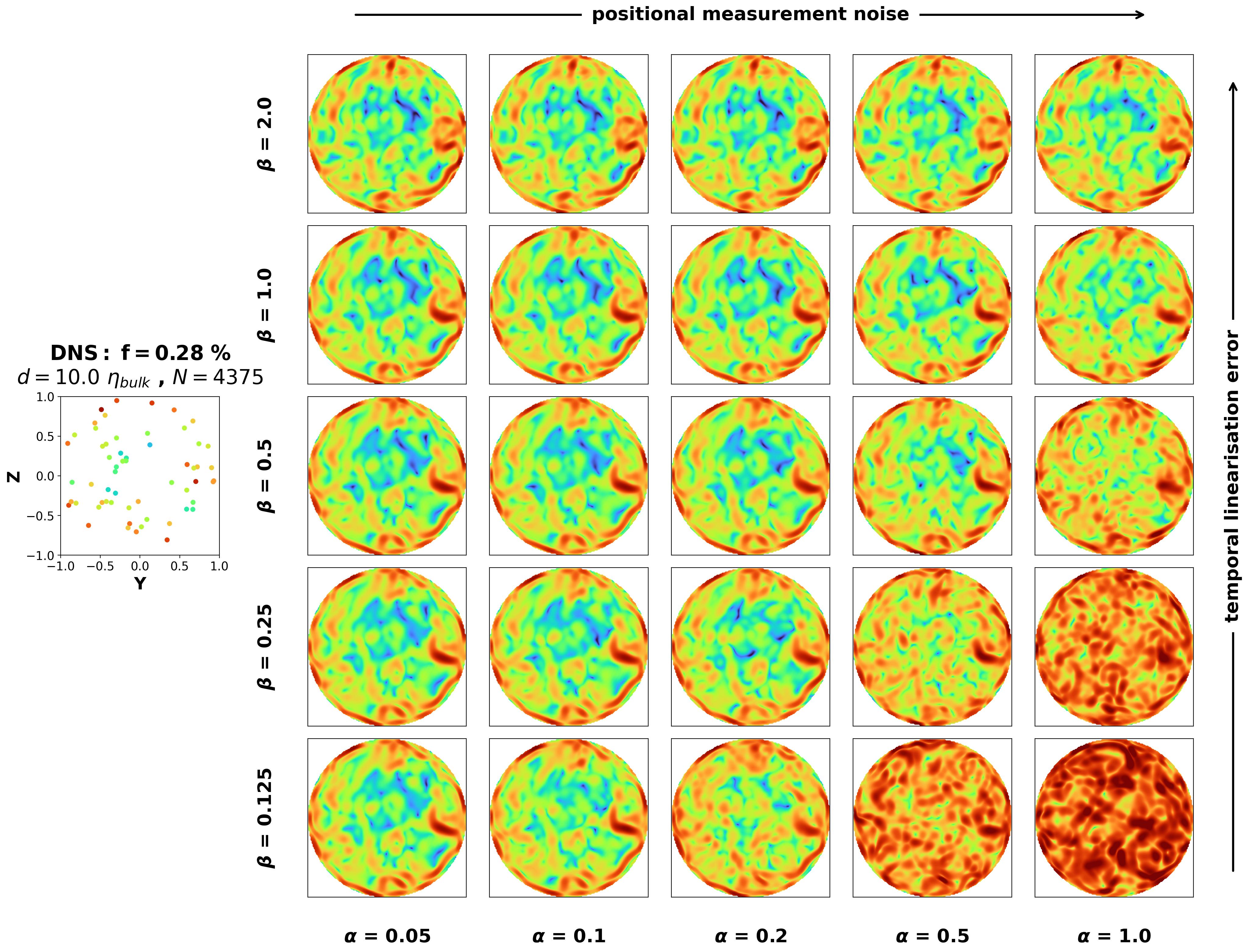}\vspace{0.2cm}
    \includegraphics[width=0.89\linewidth]{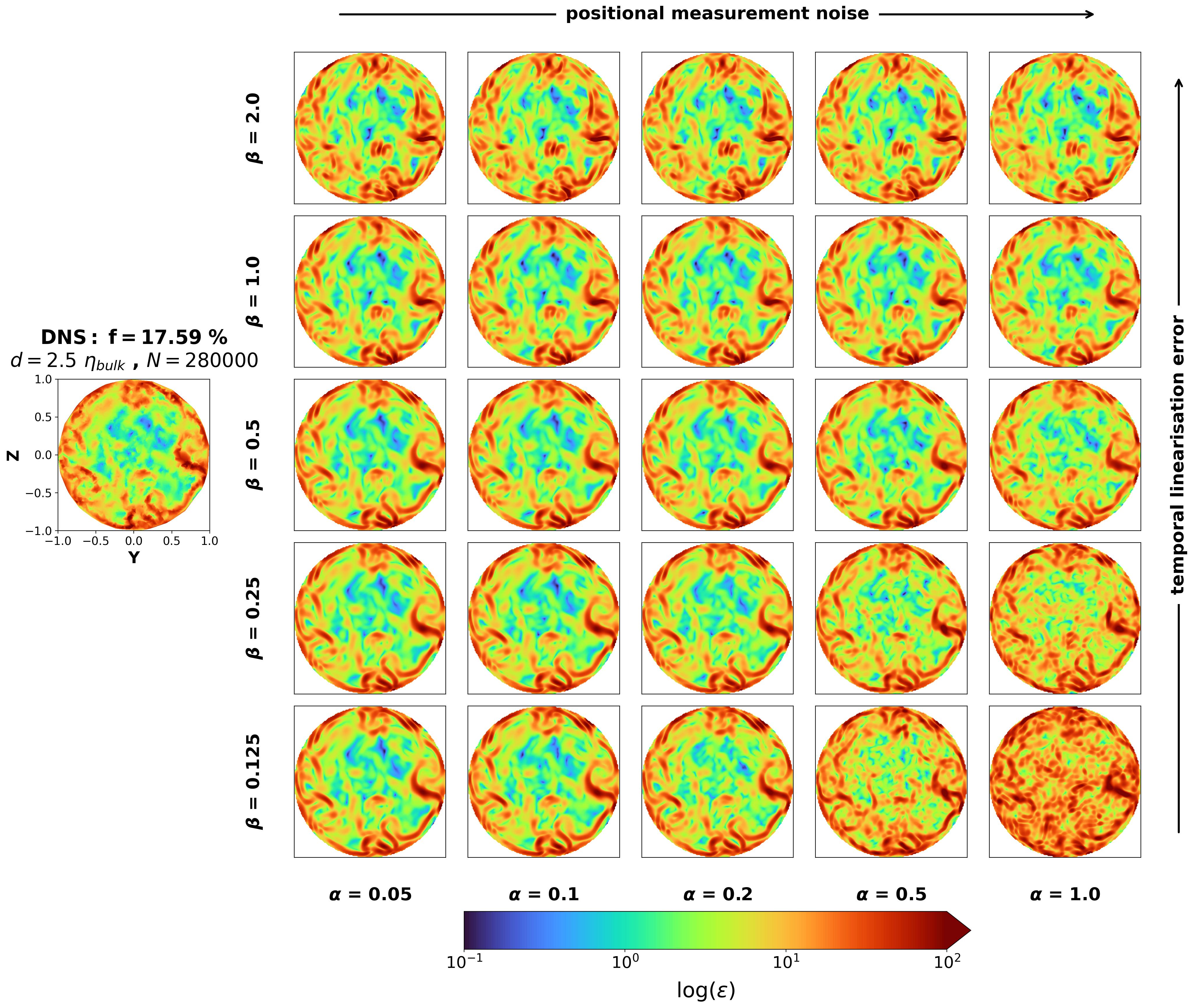}
	\caption{SnapPINN reconstructed turbulent dissipation rate $\varepsilon$ for the two sparsity datasets with $f=0.28\%$ or $N=4375$ particles (top) and $f=17.59\%$ or $N=280000$ particles (bottom) varying $\alpha$ across columns and $\beta$ across rows compared to the ground truth from the DNS (leftmost panel). The reconstructed flow fields are shown along a slice of thickness $1.5\,\eta_{\text{bulk}}$ in the pipe cross section around $X=0$.}
	\label{fig:3D_eps}
\end{figure}

\begin{figure}[H]
	\centering
    \includegraphics[width=0.89\linewidth]{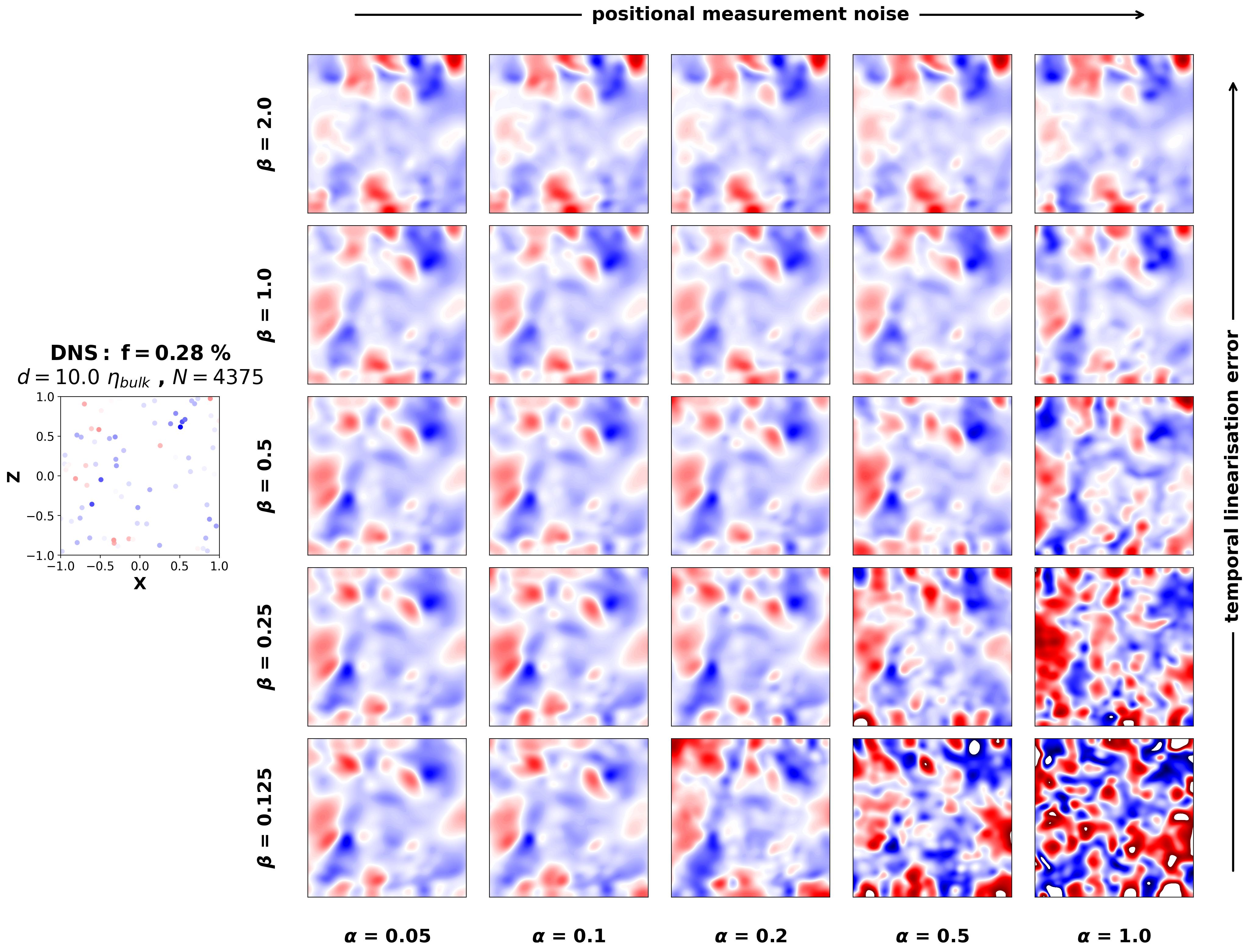}\vspace{0.1cm}
    \includegraphics[width=0.89\linewidth]{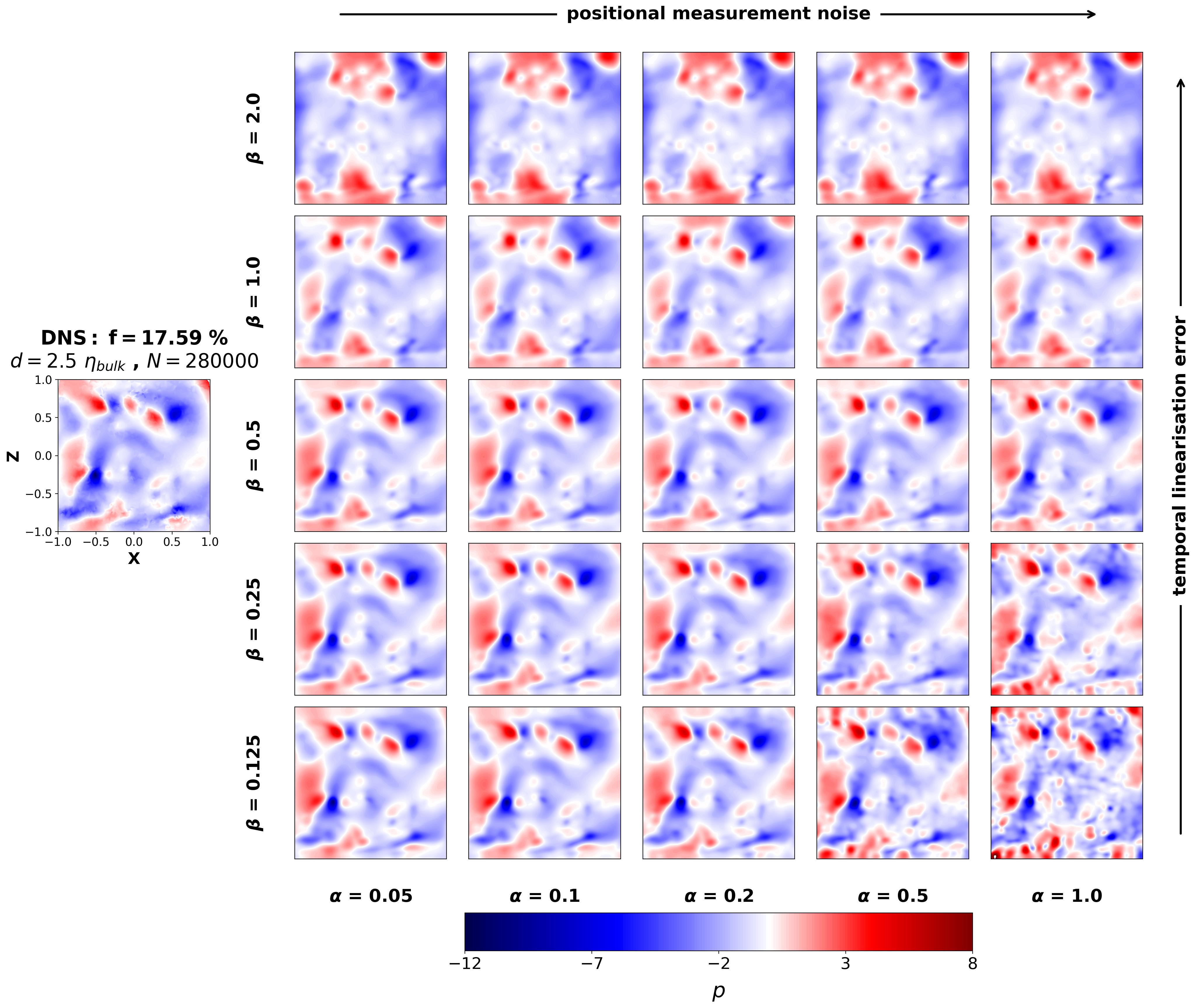}
	\caption{SnapPINN reconstructed pressure $p$ for the datasets with $f=0.28\%$ or $N=4375$ particles (top) and $f=17.59\%$ or $N=280000$ particles (bottom) varying $\alpha$ across columns and $\beta$ across rows compared to the ground truth from the DNS (leftmost panel). The reconstructed flow fields are shown along an axial slice of thickness $1.5\,\eta_{\text{bulk}}$ around $Y=0$.}
	\label{fig:3D_p}
\end{figure}

Structural agreement between the predicted flow fields and the ground truth is quantified using the PCC, and as a general rule of thumb, a PCC above $90\%$ is considered indicative of good structural agreement between the reconstructed and reference fields. To provide a qualitative interpretation of the PCC values and their relation to the reconstruction performance of SnapPINN, figures~\ref{fig:3D_eps} and \ref{fig:3D_p} present the reconstructed energy dissipation and pressure, respectively, which represent the most challenging quantities to recover due to their strong dependence on velocity gradients. To better quantify flow structures, the energy dissipation and pressure fields are shown in $1.5\,\eta_{\text{bulk}}$ thick slices at $X=0$ (cross-section) and $Y=0$ (axial-direction), respectively. Here, we focus on two sparsity cases, an extremely dilute case with $f=0.28\%$ ($N=4375$ particles and $d=10 \,\eta_{\text{bulk}}$ average interparticle spacing) and a case with moderate sparsity with $f=17.59\%$ ($N=280000$ particles and $d=2.5 \,\eta_{\text{bulk}}$ average interparticle spacing), while covering the full parameter space of $\alpha$ and $\beta$. \\
For the moderate sparsity case ($f=17.59\%$), the reconstructed energy dissipation and pressure fields closely agree with the ground truth, achieving a PCC $>90\%$ in most cases (see figure \ref{fig:pcc}), as shown in the lower panels of figures~\ref{fig:3D_eps} and \ref{fig:3D_p}. Notably, high values of $\beta$ (top rows) led to pressure reconstructions that visibly differ from the ground truth but remain physically plausible. This difference occurs because the linear discretization of the velocity represents an altered flow field compared to the exact ground truth, yet SnapPINN still successfully reconstructs a mathematically valid solution to the pressure Poisson equation for this modified field. At the other extreme—high $\alpha$ and low $\beta$ (lower-right corner)—relatively high PCC values are obtained despite noticeable noise and discrepancies in the reconstructed fields. In these instances, the high PCC reflects strong structural agreement; SnapPINN correctly localizes essential topological features, such as energy dissipation hotspots and pressure extrema, even if the overall field appears noisy. 
This observation highlights that PCC alone does not necessarily provide a complete measure of reconstruction fidelity, but still shows how SnapPINN is able to reconstruct flow fields with structural agreement to the ground truth even in cases of unreasonable high noise. \\
For the extremely dilute case ($f=0.28\%$) shown in the upper panel of figures \ref{fig:3D_eps} and \ref{fig:3D_p}, the influence of $\alpha$ and $\beta$ on the reconstruction performance of SnapPINN becomes considerably more pronounced, with only approximately one third of the cases satisfying the PCC $>90\%$ criterion for the reconstruction of energy dissipation and only one case for the reconstruction of pressure. Despite the clearly failed reconstructions near the lower-right corner of the parameter space, it is remarkable that SnapPINN retains substantial information about the spatial structure of the flow fields even at such an extremely dilute seeding situation. In particular, for many cases SnapPINN correctly identifies regions of high and low energy dissipation as well as pressure extrema, even when the detailed magnitude and spatial distribution are not fully recovered. The ability to localise these regions is particularly relevant for applications in which identifying regions of enhanced dissipation is of greater importance than recovering the exact pointwise, e.g., in the analysis of clustering hot spots in turbulent flows \citep{shaw2003particle} among many more examples. \\
Figures~\ref{fig:3D_u} and \ref{fig:3D_w} in the Appendix provide further assessment of the SnapPINN's reconstruction of the instantaneous velocities $u$ and $v$ for the two seeding density cases discussed above. The reconstruction performance follows trends similar to those for the energy dissipation rate. In particular, comparing the velocity fields in figures \ref{fig:3D_u} and \ref{fig:3D_w} at maximum noise ($\alpha=1$) with the raw, noisy training data depicted in figure \ref{fig:noise_overview} highlights the efficacy of SnapPINN's white-noise denoising capabilities. \\

\begin{figure}[H]
	\centering
    \includegraphics[width=\linewidth]{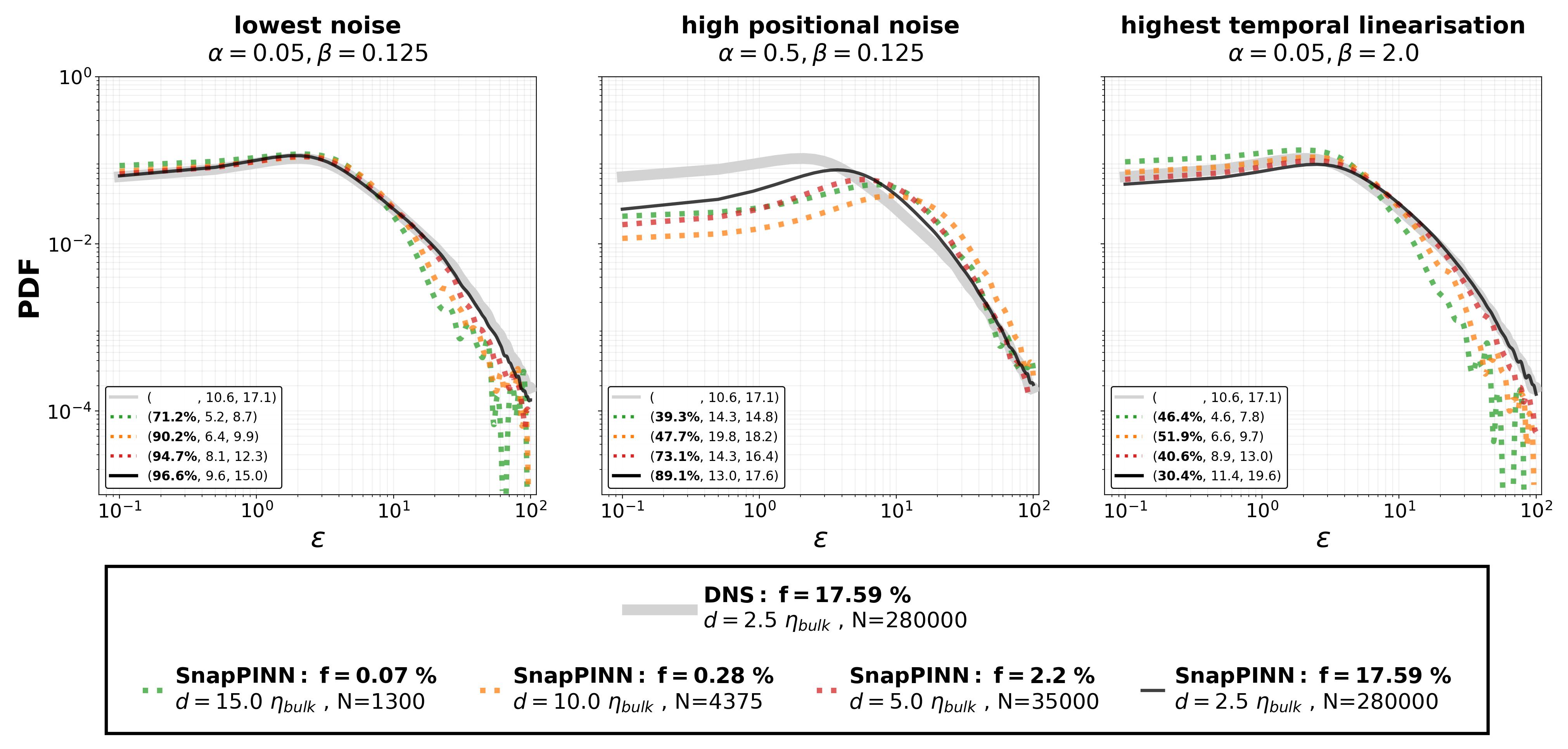}\vspace{0.1cm}
    \includegraphics[width=\linewidth]{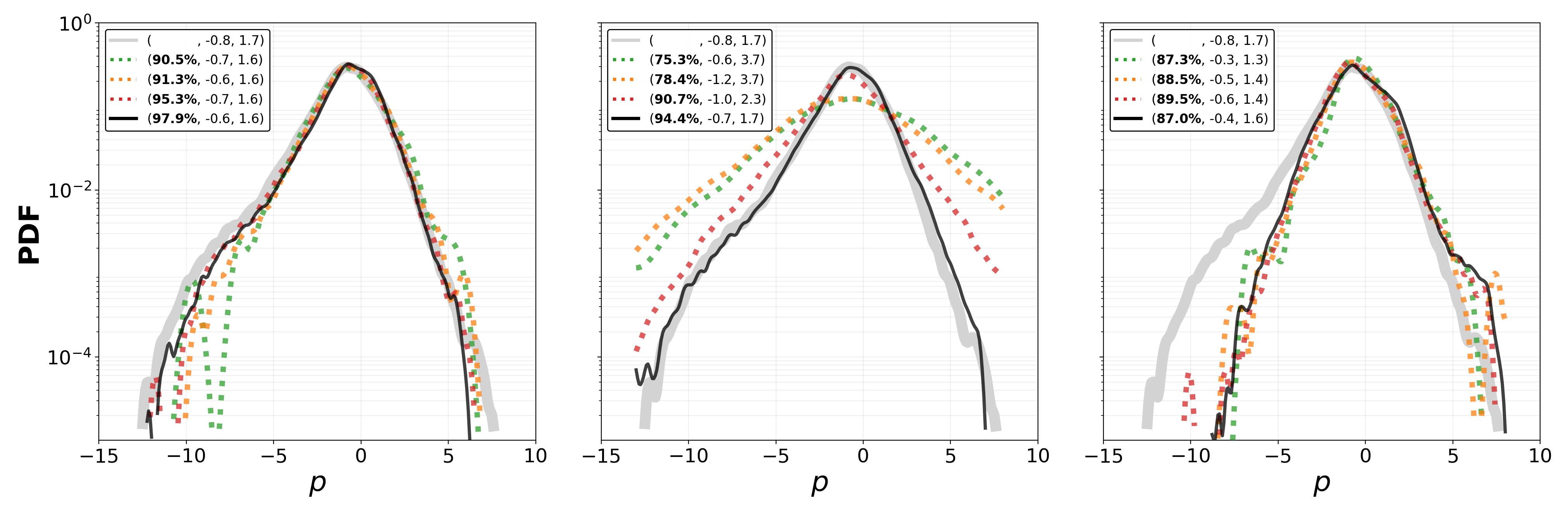}
	\caption{PDFs and the three quantities (\textbf{PCC}, mean, standard deviation) of the energy dissipation (top) and the pressure (bottom) for all sparsity cases ($f=0.07\%$, $f=0.28\%$, $f=2.2\%$, $f=17.59\%$) compared to the ground truth (thick gray line) for three selected cases: the lowest noise (left), a high positional noise (mid), and the highest temporal linearisation error (right).}
	\label{fig:pdfs3}
\end{figure}

While the PCC primarily quantifies the spatial structural agreement with the ground truth, the probability density functions (PDFs) provide a complementary assessment of the ability of SnapPINN to recover the statistical distribution and ranges of the reconstructed flow fields.
To further assess the performance of SnapPINN in characterizing instantaneous flow quantities, figure \ref{fig:pdfs3} shows the PDFs of the energy dissipation and pressure fields for selected cases from figures \ref{fig:3D_eps} and \ref{fig:3D_p}. For each seeding density, the selected cases in figure \ref{fig:pdfs3} correspond to the lowest noise case ($\alpha=0.05, \beta=0.125$), the high positional noise case ($\alpha=0.5, \beta=0.125$), and the case with the highest velocity linearisation error ($\alpha=0.05, \beta=2.0$), representing different levels of reconstruction performance. An overview of the PDFs for all cases is shown in the appendix in figure \ref{fig:pdfs1} for the reconstructed velocity components $u$ and $v$, and in figure \ref{fig:pdfs2} for the reconstructed energy dissipation and pressure fields. In figure \ref{fig:pdfs3}, the dotted coloured lines represent the PDFs for the various sparsity cases ($f=0.07\%$, $f=0.28\%$, $f=2.2\%$), while the solid black line indicates the predictions from the case with moderate sparsity ($f=17.59\%$) plotted against the DNS ground truth (thick light-gray line). Comparison of the mean and standard deviation of the reconstructed distributions with the DNS reference are detailed in the inner legends of figure \ref{fig:pdfs3}, and provides crucial insight into SnapPINN's physical fidelity. Agreement in the mean indicates that the overall magnitude of the field is captured, while agreement in the standard deviation reflects the ability to reproduce spatial variability and the intermittent nature of the turbulent flow.\\
Focusing first on the energy dissipation rate ($\varepsilon$) under the lowest noise scenario ($\alpha=0.05, \,\beta=0.125$), the DNS reference exhibits a highly skewed distribution driven by extreme intermittent events (mean = $10.6$, standard deviation = $17.1$). Under this condition with moderate sparsity ($f=17.59\%$), SnapPINN successfully reconstructs both the bulk magnitude and heavy right tail (mean = $9.6$, std = $15.0$) with a PCC of $96.6$\%. Unsurprisingly, extreme events are harder to capture from dilute data; the most dilute case ($f=0.07\%$) truncates the extreme dissipation tail, yielding a $\sim50\%$ error in mean and standard deviation (mean = $5.2$, std = $8.7$). Nevertheless, given that $\varepsilon$ spans more than four orders of magnitude and the measurements are extremely dilute, the quantitative agreement remains strong. The structural similarity is also well preserved, with the $f=0.28\%$ case still achieving a PCC above $90.2\%$.Turning next to the pressure field ($p$) under these same low-noise conditions, the DNS ground truth exhibits a much more symmetric, Gaussian-like distribution (mean = $-0.8$, std = $1.7$) but with heavy unsymmetrical tails. For this field, the reconstruction is remarkably accurate across all seeding densities. The predicted distributions closely track the DNS ground truth; even the extreme dilute configuration ($f=0.07\%$) maintains stable statistical moments (mean = $-0.7$, std = $1.6$) and achieves a PCC of $90.5$\%. The moderate sparsity case ($f=17.59\%$) accurately captures both the bulk and tails with a PCC of $97.9$
In contrast, under high positional noise ($\alpha=0.5, \,\beta=0.125$) at moderate sparsity ($f=17.59\%$), the mean of the energy dissipation is slightly overpredicted ($13.0$) but the correct spatial variability is maintained (std = $17.6$). As visually confirmed in figure~\ref{fig:3D_eps}, structural agreement remains highly robust (PCC = $89.1$\%), underscoring SnapPINN’s strong denoising capabilities despite the sensitivity of $\varepsilon$ to velocity-gradient errors. Conversely, the pressure field PDFs (bottom panel of figure \ref{fig:pdfs3}) display the largest variability because indirect inference via the Poisson pressure equation makes it highly vulnerable to accumulated velocity-gradient errors. Consequently, under high positional noise and extremely low data density ($f=0.07\%$), the pressure reconstruction fails, resulting in a flattened distribution (std = $3.7$ vs $1.7$). However, the case with moderate sparsity ($f=17.59\%$) demonstrates remarkable resilience already, closely reproducing the ground-truth field despite substantial positional noise (mean = $-0.7$ vs $-0.8$; std = $1.7$).\\
Furthermore, high temporal linearisation ($\alpha=0.05, \,\beta=2.0$) introduces distinct challenges. For energy dissipation, the reconstruction generally captures the bulk behaviour but diverges at the extremes, yielding a mean of $11.4$ and a standard deviation of $19.6$ when sparsity is moderate ($f=17.59\%$). For the bulk pressure distribution, all sparsity configurations struggle to resolve the left tail of the PDF, missing the extreme negative pressure fluctuations associated with fast-moving coherent turbulent structures. This discrepancy arises because, although the reconstruction remains physically consistent (high PCC over all sparsity cases with similar std), it no longer agrees with the DNS reference; the training velocity fields are excessively deformed due to the linear approximation.\\
It should be emphasized that Figure~\ref{fig:3D_eps} deliberately focuses on two challenging cases with high positional and temporal discretization errors to assess the robustness of SnapPINN under extreme measurement conditions. The complete parameter sweep in Figures~\ref{fig:pdfs1} and \ref{fig:pdfs2} provides a broader and more representative assessment, demonstrating robust reconstruction across the wide range of intermediate measurement conditions. For low to moderate positional noise ($\alpha \le 0.2$), predicted distributions for $\varepsilon$ and $p$ closely track the DNS ground truth across nearly all temporal resolutions (besides the highest deformation of training velocity fields at $\beta=2.0$) and most seeding densities. Severe structural deviations are strictly limited to parameter extremes (high $\alpha$ with low $\beta$, or maximum $\beta$), demonstrating the model's general robustness and broad applicability even in cases beyond reasonable experimental settings.
\section{Conclusion}
\label{Conclusion}
We introduced a novel two-stage Snapshot Physics-Informed Neural Network (SnapPINN) for reconstructing three-dimensional velocity, pressure, and turbulent dissipation fields from a single, noisy time snapshot of sparsely sampled particle data. The main outcomes of this study can be summarised as follows:
\begin{itemize}
    \item \textbf{Enabling the reconstruction of flow fields from a single velocity snapshot.} SnapPINN demonstrates that, within a broad range of experimentally relevant conditions, velocity fields, spatial gradients, and pressure can be reliably reconstructed from a single, noisy snapshot of sparse velocity data without requiring time-resolved measurements. It achieves this via a novel two-stage architecture that decouples velocity and pressure reconstruction. In Stage~1, SnapPINN acts as a physics-constrained noise filter, fitting the velocity measurements while enforcing incompressibility to recover a velocity field with physically consistent spatial gradients. Stage~2 uses these gradients to solve a single Pressure Poisson Equation. This avoids the ill-posed optimisation of unified PINNs, which struggle with instabilities when computing second-order derivatives ($\Delta \vec{u}\,/\,\text{Re}_{\tau}$) from noisy data in early-training stages using the full Navier--Stokes equations. Beyond the field reconstruction considered here, SnapPINN provides all terms required to evaluate $d\vec{u}/dt$ in equation~(\ref{eq:mom}) from a single snapshot. This opens the possibility of coupling SnapPINN with temporal solvers to forecast complex flows from highly sparse initial conditions.
    
  \item \textbf{Reliable estimation of integral flow quantities.} SnapPINN in many cases considered here accurately recovered both integral flow statistics and instantaneous flow features from a single velocity snapshot. The bulk velocity remained within $0.5\%$ of the DNS reference even at the sparsest seeding, while the more challenging gradient-based quantities showed increasing sensitivity to data sparsity. The reconstructed $\varepsilon_{\text{bulk}}$ deviated from the DNS reference by less than $10\%$ for moderately sparse cases and remained within $50\%$ even under extremely dilute conditions, representing substantial recovery for a gradient-sensitive quantity spanning more than four orders of magnitude. Furthermore, $\mathrm{Re}_{\tau}$ is recovered a~posteriori with errors ranging from approximately $4\%$ to $24\%$ at extremely dilute cases, despite never being provided as a training target, demonstrating the potential of SnapPINN for data-driven estimation of global flow parameters.
  
  \item \textbf{Reliable reconstruction of instantaneous flow features.} Spatial agreement quantified by the Pearson correlation coefficient (figure~\ref{fig:pcc}) shows that the velocity components are the most robust, while pressure and dissipation exhibit the expected degradation under increasing noise and decreasing sampling density. The probability density functions (figures~\ref{fig:pdfs1}--\ref{fig:pdfs2}) confirm that SnapPINN largely preserves the distributional shape of the reconstructed quantities, with the largest discrepancies concentrated in the tails for the most challenging cases, which occur for combinations of strong positional perturbations ($\alpha\ge 0.5$), sparse sampling, and or large velocity-estimation lags (especially $\beta=2$) that distort the underlying velocity gradients. The three-dimensional reconstructions (figures~\ref{fig:3D_eps},\ref{fig:3D_p},\ref{fig:3D_u},\ref{fig:3D_w}) further indicate that dominant coherent structures are recovered even in extremely dilute cases, whereas fine-scale fluctuations and high-gradient regions remain the most sensitive to measurement noise. 

\item \textbf{Practical reliability map for applications without ground truth.} The systematic variation of data sparsity, positional uncertainty, and temporal discretization across 100 cases is consolidated into a quantitative reliability map (Figure~\ref{fig:pcc}) linking experimental measurement conditions to expected reconstruction accuracy with SnapPINN. The experimentally accessible parameters $\alpha$, $\beta$, and $d$ can then be used to locate a given measurement configuration within this map and assess which reconstructed flow quantities can be reliably interpreted when ground-truth data are unavailable. Although derived from turbulent pipe flow, its strong near-wall gradients, anisotropy, and broad dynamic range across velocity components provide a particularly challenging test case, and the resulting map may therefore serve as a useful first-order guideline for other turbulent flow configurations.

\end{itemize}
\section*{Appendix}
\setcounter{figure}{0}

\renewcommand{\thefigure}{A\arabic{figure}}
\begin{figure}[H]
	\centering
    \includegraphics[width=0.85\linewidth]{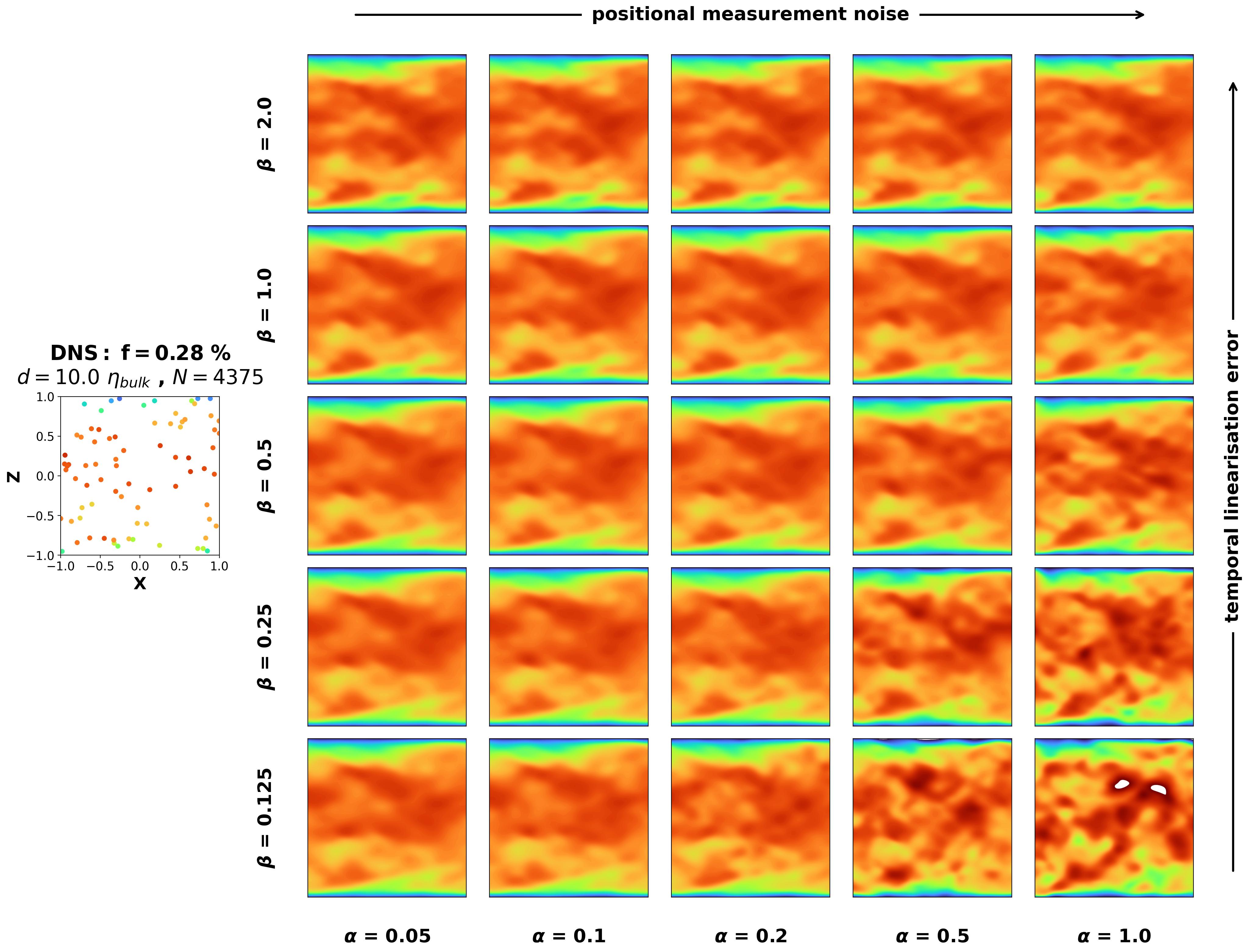}\vspace{0.2cm}
    \includegraphics[width=0.85\linewidth]{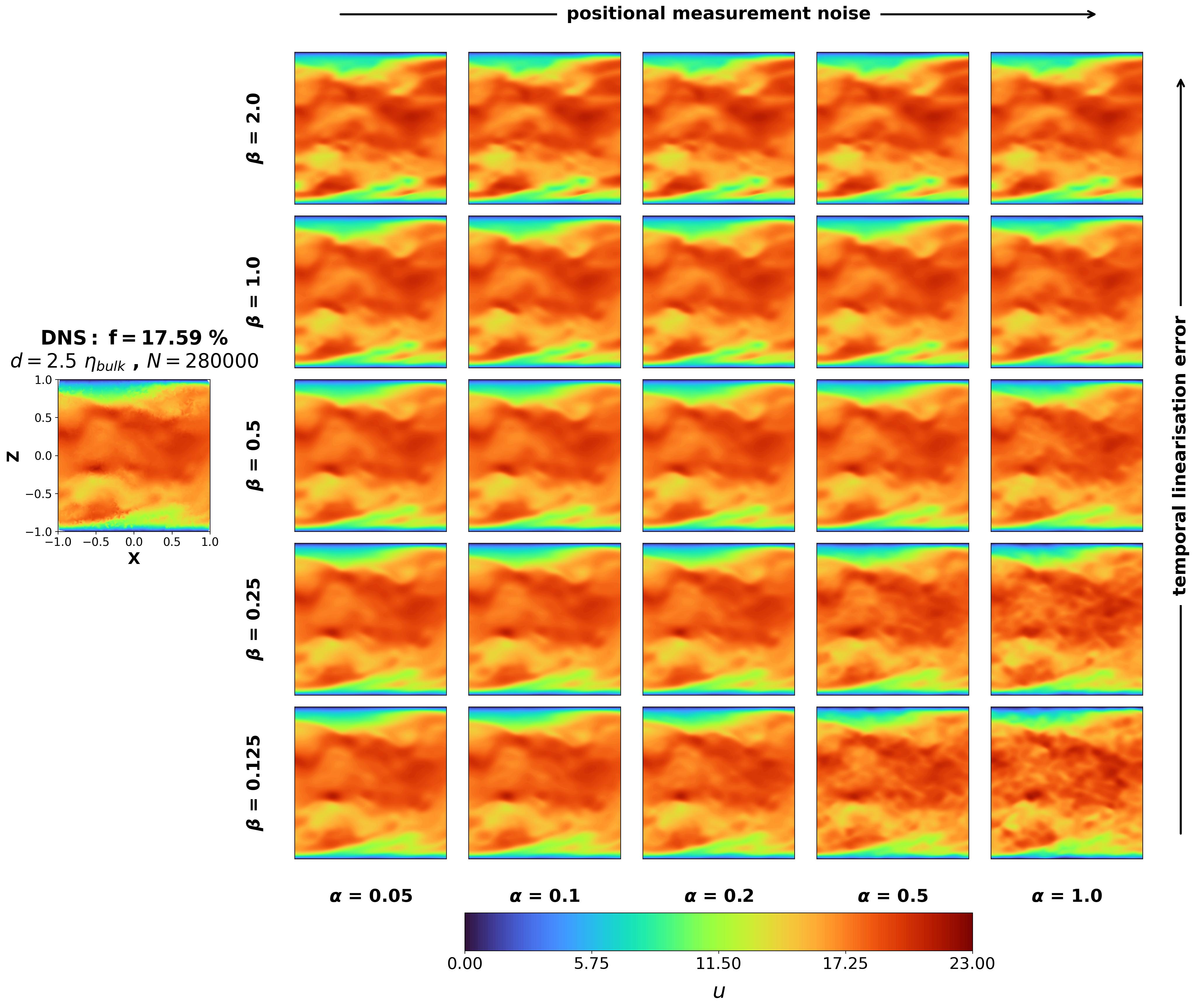}
	\caption{SnapPINN reconstructed velocity component $u$ for the datasets with $f=0.28\%$ or $N=4375$ particles (top) and $f=17.59\%$ or $N=280000$ particles (bottom) varying $\alpha$ across columns and $\beta$ across rows compared to the ground truth from the DNS (leftmost panel). The reconstructed flow fields are shown along an axial slice of thickness $1.5\,\eta_{\text{bulk}}$ around $Y=0$.}
	\label{fig:3D_u}
\end{figure}
\begin{figure}[H]
	\centering
    \includegraphics[width=0.89\linewidth]{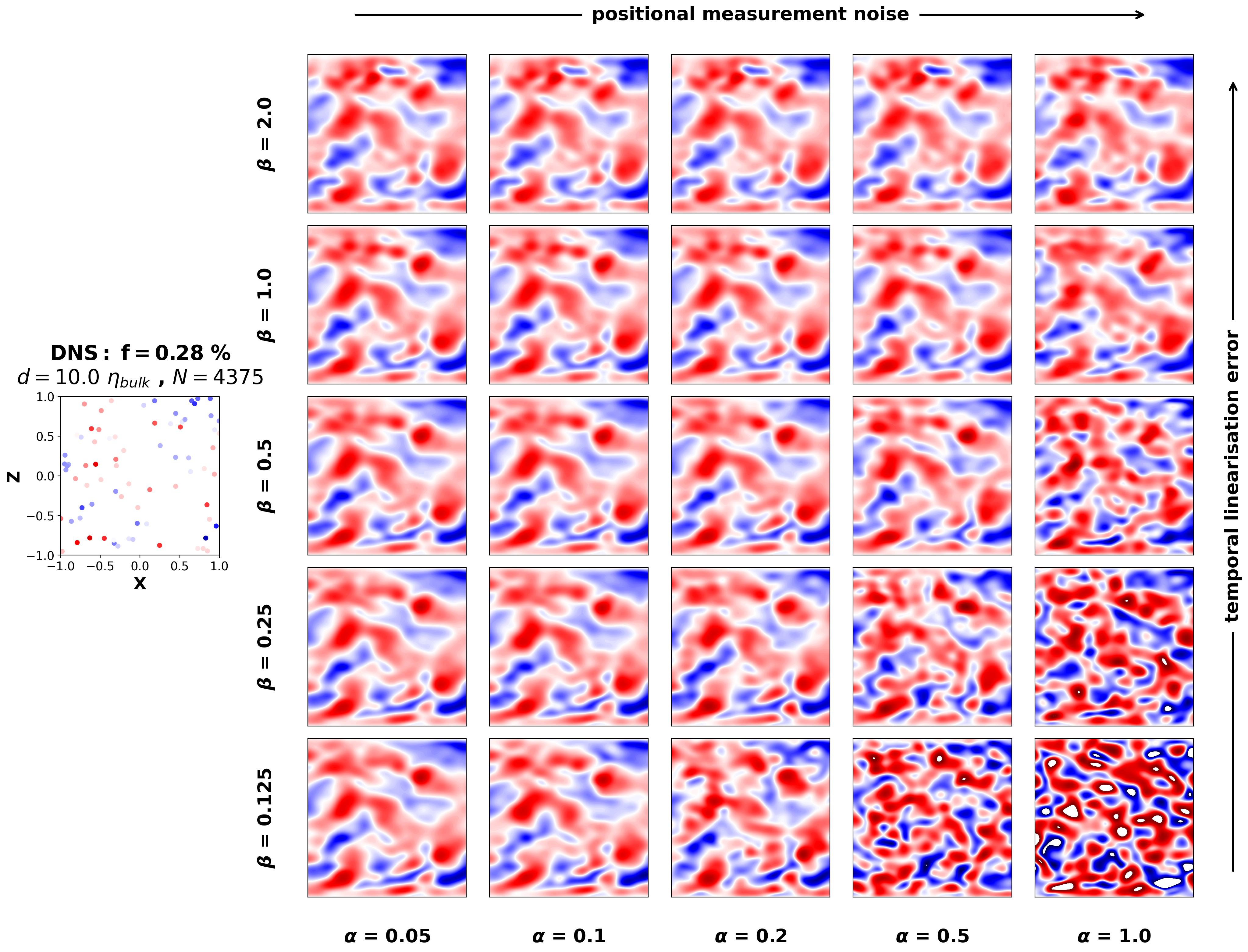}\vspace{0.2cm}
    \includegraphics[width=0.89\linewidth]{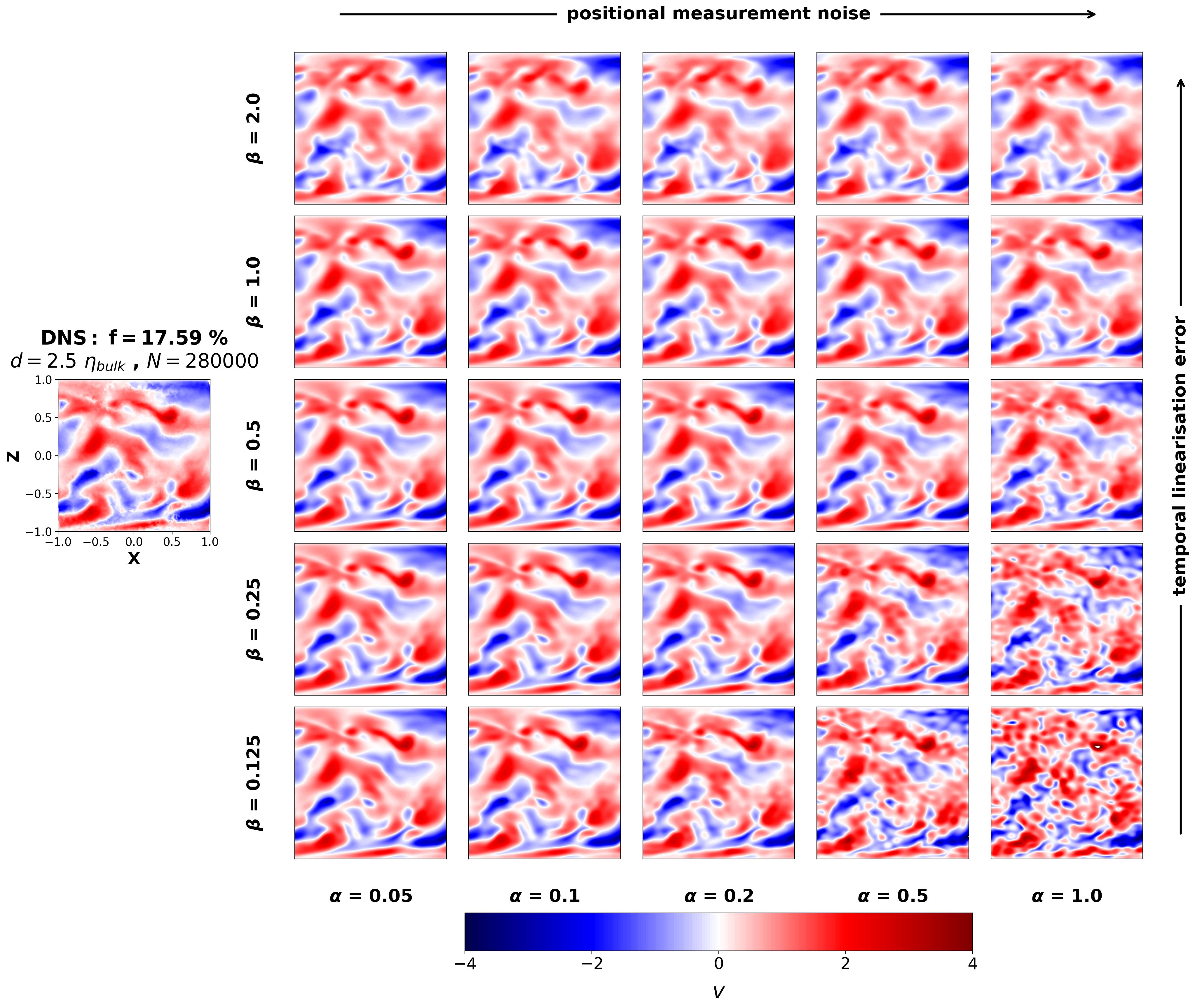}
	\caption{SnapPINN reconstructed velocity component $v$ for the datasets with $f=0.28\%$ or $N=4375$ particles (top) and $f=17.59\%$ or $N=280000$ particles (bottom) varying $\alpha$ across columns and $\beta$ across rows compared to the ground truth from the DNS (leftmost panel). The reconstructed flow fields are shown along an axial slice of thickness $1.5\,\eta_{\text{bulk}}$ around $Y=0$.}
	\label{fig:3D_w}
\end{figure}
\begin{figure}[H]
	\centering
    \includegraphics[width=0.74\linewidth]{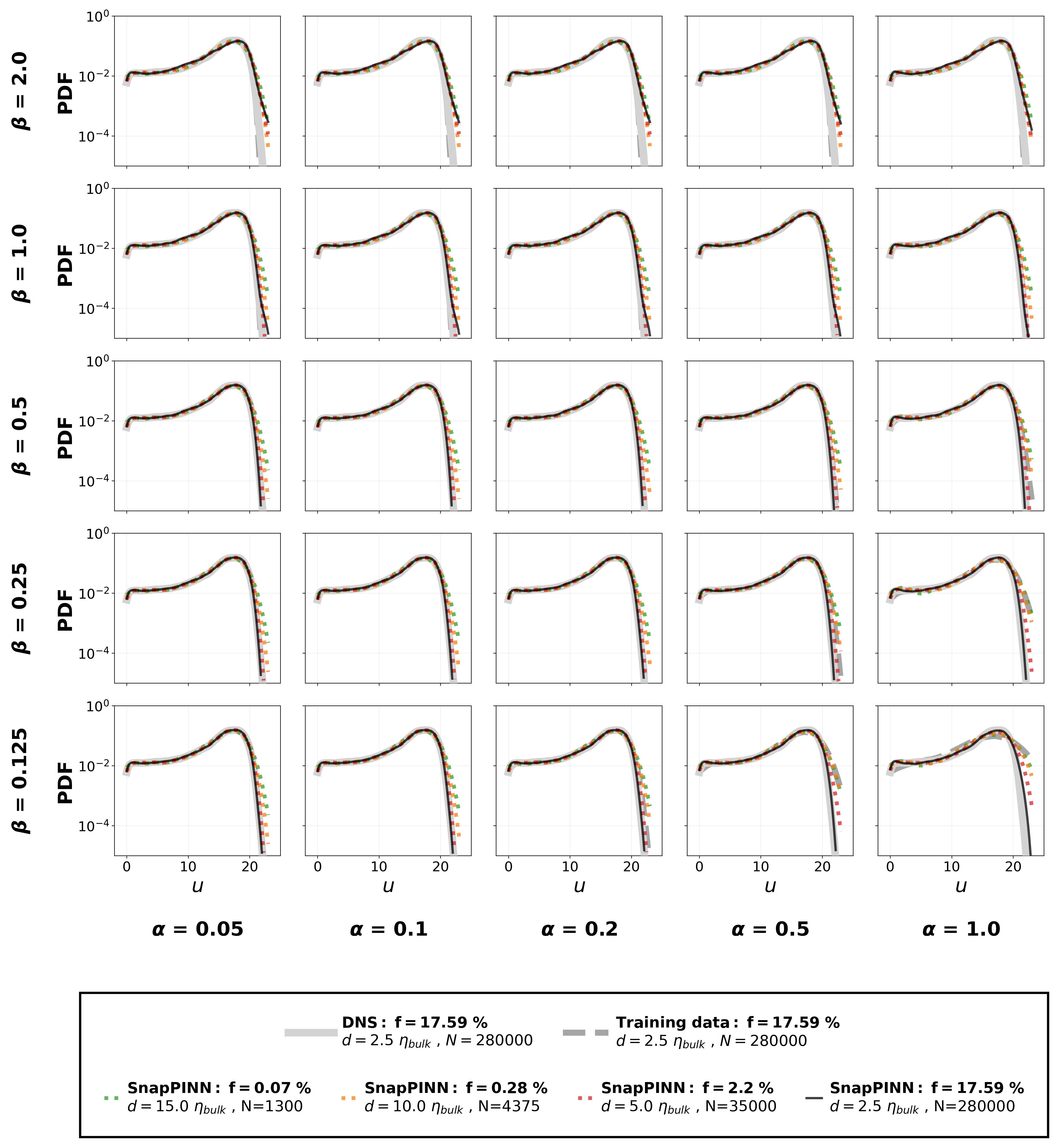}\vspace{0.23cm}
    \includegraphics[width=0.74\linewidth]{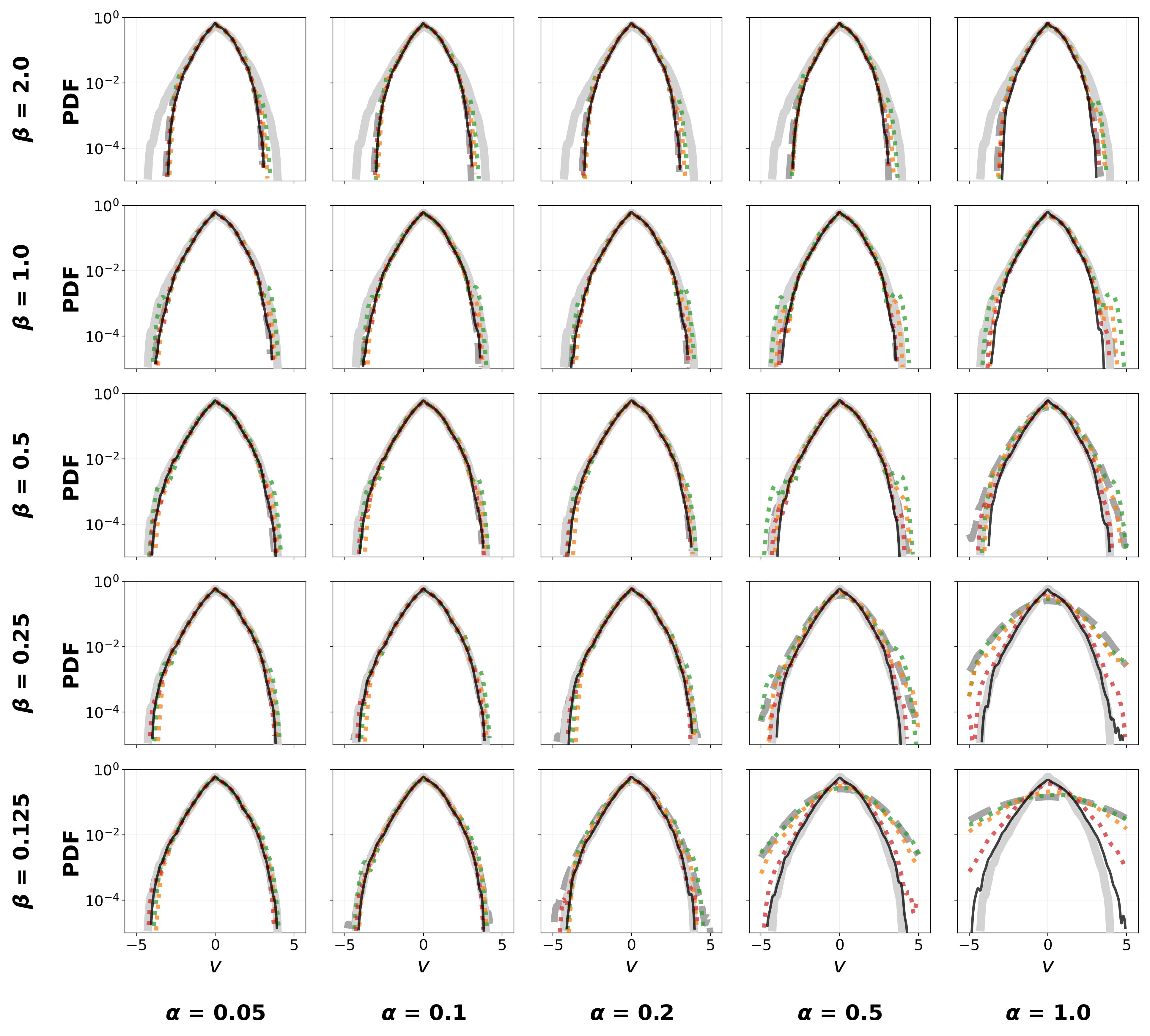}
	\caption{PDFs of reconstructed velocity components, $u$ (top), and $v$ (bottom) using SnapPINN for all sparsity cases ($f = 0.07$\%, $f = 0.28$\%, $f = 2.2$\%, $f = 17.59$\%) compared to the ground truth (thick gray line).}
	\label{fig:pdfs1}
\end{figure}
\begin{figure}[H]
	\centering
    \includegraphics[width=0.74\linewidth]{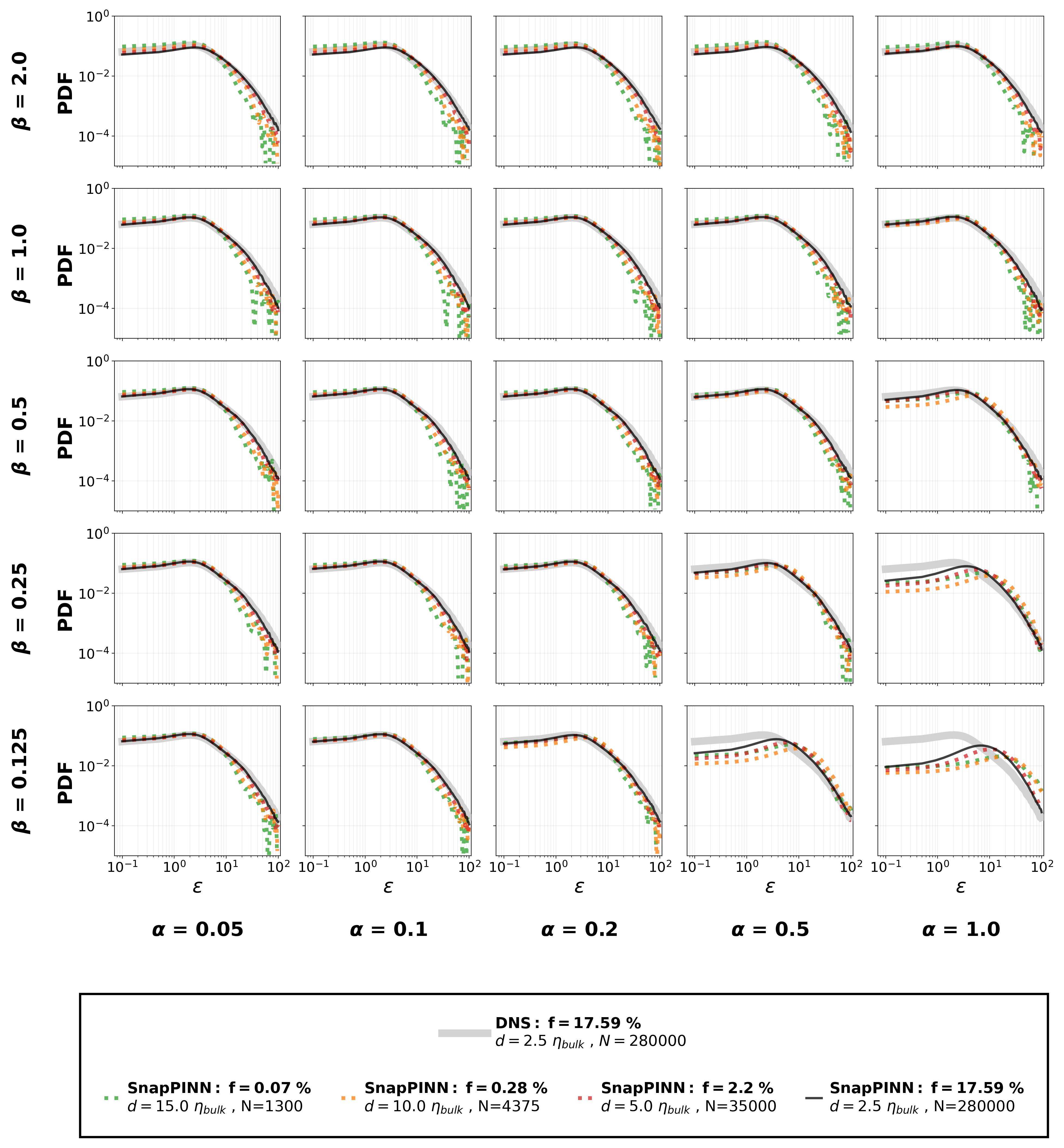}\vspace{0.23cm}
    \includegraphics[width=0.74\linewidth]{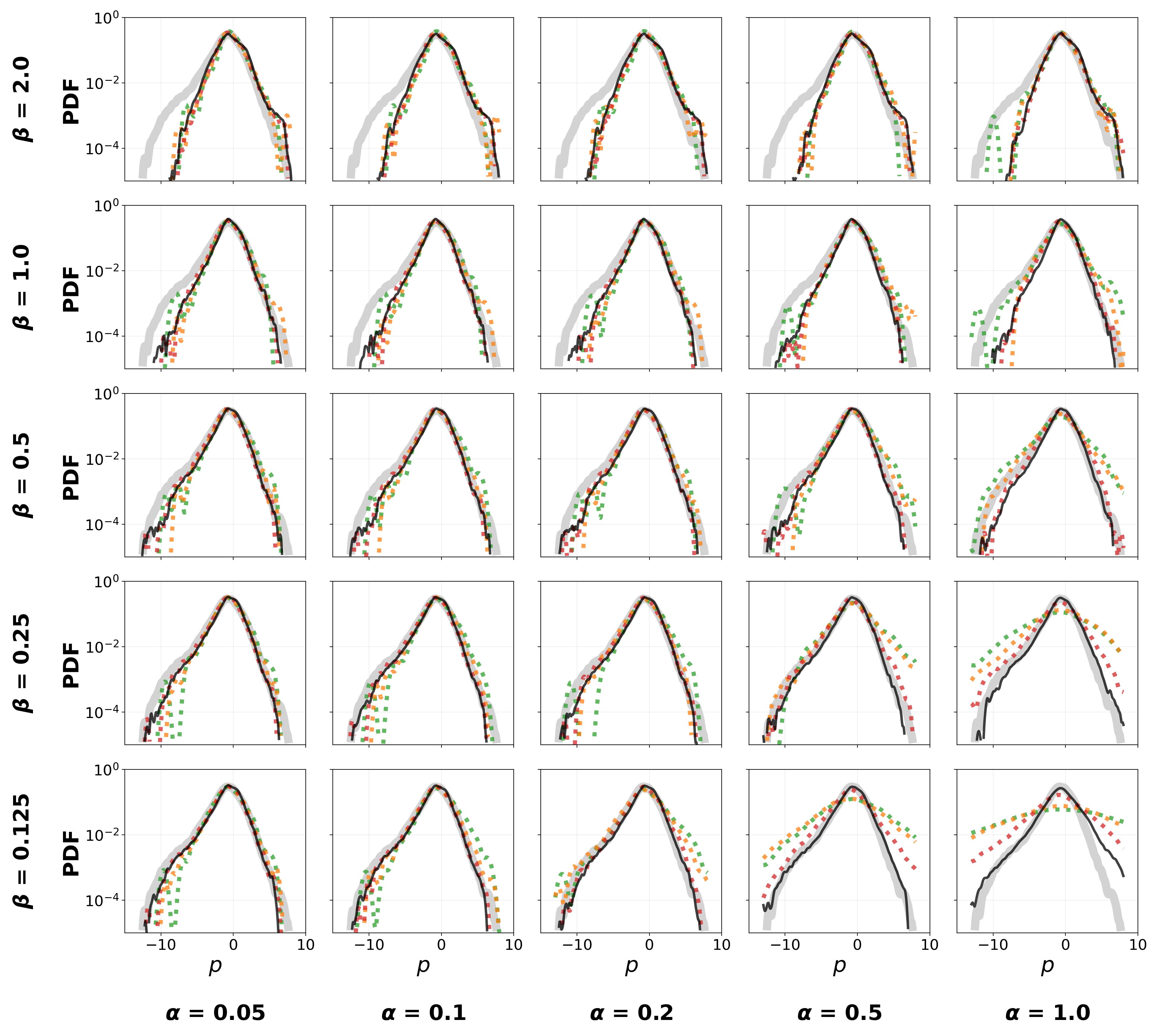}
	\caption{PDFs of reconstructed energy dissipation $\varepsilon$ (top), and the pressure $p$ (bottom) using SnapPINN for all sparsity cases ($f = 0.07$\%, $f = 0.28$\%, $f = 2.2$\%, $f = 17.59$\%) compared to the ground truth (thick gray line).}
	\label{fig:pdfs2}
\end{figure}

\subsection*{Acknowledgments}
\noindent The authors gratefully acknowledge the scientific support and HPC resources provided by the German Aerospace Center (DLR). The HPC system CARA is partially funded by "Saxon State Ministry of Economic Affairs, Labour, Energy and Climate Action" and " Federal Ministry of Research, Technology and Space". The HPC system CARO is partially funded by "Ministry of Science and Culture of Lower Saxony" and " Federal Ministry of Research, Technology and Space".

\subsection*{Declaration of Interests}
The authors report no conflict of interest.

\bibliographystyle{plainnat}
\bibliography{refs}

\end{document}